\documentclass[ba,,preprint]{imsart}
\pdfoutput=1 
\pubyear{2021}
\volume{TBA}
\issue{TBA}
\firstpage{1}
\lastpage{1}

\usepackage{amsthm}
\usepackage{amsmath}
\usepackage{natbib}
\usepackage[colorlinks,citecolor=blue,urlcolor=blue,filecolor=blue,backref=page]{hyperref}
\usepackage{graphicx}

\startlocaldefs
\numberwithin{equation}{section}
\theoremstyle{plain}

\usepackage{microtype}
\usepackage{xcolor}
\usepackage{array}
\newcolumntype{H}{>{\setbox0=\hbox\bgroup}c<{\egroup}@{}} 
\usepackage{makecell} 
\endlocaldefs

\begin{document}


\begin{frontmatter}
\title{Prior knowledge elicitation: The past, present, and future}
\runtitle{}

\begin{aug}
\author{\fnms{Petrus} \snm{Mikkola}\thanksref{addr1}\ead[label=e1]{petrus.mikkola@aalto.fi}},
\author{\fnms{Osvaldo A.} \snm{Martin}\thanksref{addr1,addr6}\ead[label=e2]{osvaldo.martin@aalto.fi}},
\author{\fnms{Suyog} \snm{Chandramouli}\thanksref{addr1,addr2}\ead[label=e3]{suyog.chandramouli@aalto.fi}},
\author{\fnms{Marcelo} \snm{Hartmann}\thanksref{addr2}\ead[label=e4]{marcelo.hartmann@helsinki.fi}},
\author{\fnms{Oriol} \snm{Abril Pla}\thanksref{addr2}\ead[label=e5]{oriol.abrilpla@helsinki.fi}},
\author{\fnms{Owen} \snm{Thomas}\thanksref{addr3}\ead[label=e6]{o.m.t.thomas@medisin.uio.no}},
\author{\fnms{Henri} \snm{Pesonen}\thanksref{addr3}\ead[label=e7]{h.e.pesonen@medisin.uio.no}},
\author{\fnms{Jukka} \snm{Corander}\thanksref{addr3,addr2,addr7}\ead[label=e8]{jukka.corander@medisin.uio.no}},
\author{\fnms{Aki} \snm{Vehtari}\thanksref{addr1}\ead[label=e9]{aki.vehtari@aalto.fi}},
\author{\fnms{Samuel} \snm{Kaski}\thanksref{addr1,addr5,equalcont}\ead[label=e10]{samuel.kaski@aalto.fi}},
\author{\fnms{Paul-Christian} \snm{Bürkner}\thanksref{addr4,equalcont}\ead[label=e11]{paul-christian.buerkner@simtech.uni-stuttgart.de}}
\and
\author{\fnms{Arto} \snm{Klami}\thanksref{addr2,equalcont,correspon}\ead[label=e12]{arto.klami@helsinki.fi}}


\address[addr1]{Helsinki Institute of Information Technology, Department of Computer Science (PM,OAM,AK,SK) \textcolor{white}{XX} and Department of Information and Communication Engineering (SC), Aalto University, Finland}

\address[addr2]{Helsinki Institute of Information Technology, Department of Computer Science (MH,OA,AK,SC) \textcolor{white}{XX} and Department of Mathematics and Statistics (JC), University of Helsinki, Finland    
}

\address[addr3]{Institute of Basic Medical Sciences, University of Oslo (JC,HP) and\\  \textcolor{white}{XX} Akershus Universitetssykehus (OT), Norway
}

\address[addr4]{Cluster of
Excellence SimTech, University of Stuttgart, Germany
}

\address[addr5]{Department of Computer Science, University of Manchester, UK}

\address[addr6]{Instituto de Matemática Aplicada San Luis, CONICET-UNSL, Argentina}

\address[addr7]{Parasites and Microbes, Wellcome Sanger Institute, UK}

\address[equalcont]{Equal contribution}
\address[correspon]{Corresponding author, \printead{e12}}

\end{aug}

\begin{abstract}
    Specification of the prior distribution for a Bayesian model is a central part of the Bayesian workflow for data analysis, but it is often difficult even for statistical experts. In principle, prior elicitation transforms domain knowledge of various kinds into well-defined prior distributions, and offers a solution to the prior specification problem. In practice, however, we are still fairly far from having usable prior elicitation tools that could significantly influence the way we build probabilistic models in academia and industry. We lack elicitation methods that integrate well into the Bayesian workflow and perform elicitation efficiently in terms of costs of time and effort. We even lack a comprehensive theoretical framework for understanding different facets of the prior elicitation problem.

    Why are we not widely using prior elicitation? We analyse the state of the art by identifying a range of key aspects of prior knowledge elicitation, from properties of the modelling task and the nature of the priors to the form of interaction with the expert. The existing prior elicitation literature is reviewed and categorized in these terms. This allows recognizing under-studied directions in prior elicitation research, finally leading to a proposal of several new avenues to improve prior elicitation methodology.
\end{abstract}


\end{frontmatter}


\section{Introduction}

Bayesian statistics uses probabilistic models, formalized as a set of interconnected random variables following some assumed probability distributions, for describing observations. Designing a suitable model for a given data analysis task requires both significant statistical expertise and domain knowledge, and is typically carried out as an iterative process that involves repeated testing and refinement. This process can be formulated as the \emph{Bayesian workflow} to aid the modeller work in a more reproducible and documentable manner; see \citet{Gelman2020} for a recent detailed formalization partitioning the process into numerous sub-workflows focusing on different facets of the process, such as model specification, inference and model validation.

We focus on one central part of that Bayesian workflow: the choice of prior distributions for the parameters of the model. In particular, we discuss approaches to \emph{eliciting} knowledge from a domain expert to be converted into prior distributions suitable for use in a probabilistic model, rather than assuming the analyst can specify the priors directly. The fundamental goal of this \emph{expert knowledge or prior elicitation} process (defined in Section~\ref{What_is_prior_elicitation}) is to help practitioners design models that better capture the essential properties of the system or process under study. Good elicitation tools could also help in the additional goal of fostering wide-spread adoption of probabilistic modelling by reducing the required statistical expertise. An ideal prior elicitation approach would simultaneously make model specification faster, easier, and better at representing the knowledge of the expert. It is hoped that availability of good prior elicitation tools would qualitatively transform the process of prior specification within the Bayesian modelling workflow, analogously to what \emph{probabilistic programming} languages and their efficient model-agnostic algorithms have done for model specification and inference \citep[e.g.][]{stan, pymc, turing}.

Prior elicitation has a long history dating back to the 1960s \citep{winkler:1967}, and excellent textbook accounts \citep{ohagan:2006}, surveys and reviews \citep{garthwaite:2005,ohagan:2019} are available. Despite the established problem formulation and broad scientific literature on methods for eliciting priors in different special cases -- often for some particular model family -- we are still lacking practical tools that would routinely be used as part of the modelling workflow. While a few actively developed tools for interactive prior elicitation exist and are used in selected domains, exemplified by \texttt{SHELF} \citep{oakleyshelf:2019} and \texttt{makemyprior} \citep{hem2021}, their active user-base remains a tiny fraction of people regularly applying probabilistic models. Instead, practitioners often use rather ad hoc procedures to specify and modify the priors \citep[e.g.][]{sarma:2020}, building on personal expertise and experience, ideally learned by following literature on prior recommendations -- for instance by \citet{stan}, on logistic regression \citep{gelman:2008,ghosh:2018}, on hierarchical models \citep{gelman2006prior,simpson:2017,chung:2015}, on Gaussian random fields \citep{fuglstad:2019}, or on autoregressive processes \citep{sigrunn:2017}.

We discuss reasons for the still limited impact of prior elicitation research on prior specification in practice, and propose a range of research directions that need to be pursued to change the situation. Our main claim is that we are still fairly far from having practical prior elicitation tools that could significantly influence the way probabilistic models are built in academia and industry. To improve over the current state, coordinated research involving expertise from multiple disciplines is needed. This paper is both our call for experts to join these efforts, and a concrete guide for future research. Consequently, the paper is written both for people already developing prior elicitation techniques and for people working on specific complementary problems, who we are encouraging to contribute to the common goal. For people looking for practical methods for prior elicitation in their own modelling problems, we unfortunately cannot yet provide very concrete solutions, but we are looking for your feedback on the requirements and desired goals.

As will be clarified later, several interconnected elements hinder the uptake of prior elicitation methods. Some of these are purely \emph{technical} properties of the elicitation algorithms, relating to limited scope in terms of models that prevents their use in general probabilistic programming, or ability to only address univariate priors, sequentially, rather than jointly eliciting all priors of a model. Some are more \emph{practical}, such as many of the approaches still being too difficult for non-statistical experts to use, and lack of good open source software that integrates well with the current probabilistic programming tools used for other parts of the modelling workflow. Finally, some aspects are more \emph{societal}: The concrete value of prior elicitation has not yet been adequately demonstrated in highly visible case studies, and hence end-users do not know to request better approaches, and decision-makers have not invested resources for them.

Critically, these issues are highly interconnected. For building large-scale demonstrations of the practical value of prior elicitation in visible applications, we would already need to have high-quality software that integrates with existing modelling workflows, as well as elicitation methods capable of efficiently eliciting priors for models of sufficient complexity. Given that the field is currently falling short of achieving any of these aspects, we argue that significant coordinated effort is needed before we can make concrete recommendations on best practices for elicitation in any given instance. We can largely work in parallel towards mitigating these issues, but it is important to do this in a coordinated manner, typically so that researchers with complementary scientific expertise work together to address the most closely connected elements. For instance, an ideal team for designing the software tools would combine at least computer engineers, statisticians, interface designers and cognitive scientists, to guarantee that the most important aspects for all dimensions are accounted for.

To proceed towards practical recommendations, we start by identifying seven key dimensions that characterize the prior elicitation challenge and possible solutions for it, to provide a coherent framework for discussing the matter. We inspect prior elicitation from the perspectives of (1) properties of the prior distribution itself, (2) the model family and the prior elicitation method's dependence on it, (3) the underlying elicitation space, (4) how the method interprets the information provided by the expert, (5) computation, (6) the form and quantity of interaction with the expert(s), and (7) the assumed capability of the expert, both in terms of their domain knowledge and statistical understanding. We discuss all of these fundamental dimensions in detail (Section~\ref{prior_hypercube}), identifying several practical guidelines on how specific characteristics for each of them influence the desired properties for the elicitation method. We also provide a review of existing elicitation methods to highlight gaps in the available literature, but for more comprehensive reviews at earlier stages of the literature, we recommend consulting \citet{ohagan:2006} and \citet{garthwaite:2005}.

Building on this framework, we proceed to make recommendations for future research, by characterizing in more detail the current blockers listed above, and outlining our current suggestions on what kind of research is needed to resolve the issues. These recommendations are necessarily on a relatively high abstraction level, but we hope they still provide a tangible starting point for people coming from outside the current prior elicitation research community. In particular, we discuss easy-to-use software that integrates with open probabilistic programming platforms as a necessary requirement for practical impact, already outlining a possible architecture and key components for such a system. We emphasize the need for considerably extended user evaluation for verifying that the methods have practical value.

\section{Prior Elicitation}

\subsection{What is prior elicitation?}\label{What_is_prior_elicitation}

Specifying prior probability distributions over variables of interest (such as model's parameters) is an essential part of Bayesian inference. These distributions represent available information regarding values of the variables prior to considering the current data at hand. \textit{Prior elicitation} is one way to specify priors and refers to the process of eliciting the subjective knowledge of domain experts in a structured manner and expressing this knowledge as prior probability distributions \citep{garthwaite:2005,ohagan:2006}. This involves not only actually gathering the information from an expert, but also any computational methods that may be needed to transform the collected information into well-defined prior probability distributions. 

While prior elicitation is the focus of our article, it is only one of many ways to specify informative priors. Alternatively, analysts may directly specify priors based on a variety of other information sources including relevant literature or databases when the parameters have fairly concrete real-world referents \citep{gelman2013philosophy}. For instance, in medicine, data-based priors have been widely adopted
\citep{frantisek:2021}, while there are situations where prior elicitation is preferred, such as with parameter settings that are unverifiable from the data to hand \citep{dallow:2018}. When historical data are available, priors can be specified by `borrowing' from that data, known as \textit{historical borrowing} \citep{viele2014use}, using hierarchical modelling \citep{pocock1976combination,spiegelhalter2004bayesian,neuenschwander2010summarizing,neuenschwander2016use,hobbs2011hierarchical,schmidli2014robust} or through power priors \citep{ibrahim:2000,ibrahim:2015,psioda2019bayesian}.

Besides encoding domain knowledge, there are other grounds for specifying priors.
For instance, priors can be chosen such that they affect the information in the likelihood as weakly as possible (\textit{noninformative priors}), yield smoother and more stable inferences (\textit{regularizing priors}), or yield `asymptotically acceptable' posterior inference (\textit{reference priors}) \citep{Gelman2017,kass:1996}. In particular, one may require a prior to ensure posterior consistency  \citep{rousseau2016frequentist, moreno2015posterior}. While we acknowledge the validity of these approaches as well, we do not discuss them in more detail in this article due to our specific goal of investigating the state of prior elicitation, not prior specification in general. However, we repeat the general observation that practically flat priors, such as $\textrm{normal}(0,10^{6})$, sometimes used by practitioners should be avoided, due to problems in posterior inference \citep{carlin:2000,VANDONGEN:2006,gelman2006prior,BDA:2013,Gelman2017,gelman2020holes,smid2020dangers}. 

Most parameters of theory-driven, physics-based models have a clear meaning outside the model itself. For example, the weight of a star has meaning outside the statistical model that is used for weight estimation from astronomical data. However, for less precise theories and corresponding models, say, in the social sciences, parameters often only have meaning within the context of the model they are part of \citep{Gelman2017}. Accordingly, for the latter kind of models, prior elicitation procedures need to take into account that the distribution being elicited is part of a model and cannot simply be viewed in isolation. The Bayes rule connects the prior $p(\theta)$ to the posterior $p(\theta|y)$ within the context of the likelihood $p(y|\theta)$, 
\begin{equation}\label{Bayes_rule}
    p(\theta | y) = \frac{p(y|\theta)p(\theta)}{p(y)},
\end{equation}
where the observables and the parameters are denoted by $y$ and $\theta$, respectively. The goal of prior elicitation is to elicit $p(\theta)$ from an expert. In line with \citet{Gelman2017}, we note that the likelihood $p(y|\theta)$ partially determines the scale and the range of reasonable values for $\theta$. In that respect, prior elicitation differs from the elicitation for evidence-based decision-making \citep[e.g.][]{kennedy:2008,browstein:2019} or expert systems \citep[e.g.][]{studer:1998,wilson:2005}, where the objective is to elicit a probability distribution (not paired with any likelihood) that represents uncertainty on the parameters of a decision model \citep{grigore2016comparison} or the node probability tables of a Bayesian network \citep{nunes2018issues}. We note, however, that whether prior elicitation should depend on the (sampling) model is still under community debate and there is no universally accepted answer yet.

A common elicitation process involves two persons, called expert and analyst. 
We follow the convention that the expert is referred to as a female and the analyst as a male \citep{ohagan:2007}, and use the term \textit{analyst} instead of facilitator to emphasize that the analyst can play many roles simultaneously \citep[][Section 2.2.1]{ohagan:2006}, for instance, as a statistician and a facilitator. The \textit{facilitator} is an expert in the process of elicitation. He can take an active role such as manage dialogue between the expert(s) or a more passive role such as assisting in the elicitation between the expert and an elicitation algorithm. Not all elicitation methods require a human facilitator, but instead he/it is built into the elicitation software \cite[see an interesting alternative definition by][]{kahle:2016}. The \textit{expert} refers to the domain expert, who is also called a substantive expert. She has relevant knowledge about the uncertain quantities of interest, such as the model parameters or observables. For more about the definition and recruitment of the experts, see \citet{bolger:2018}.

\subsection{Why isn't the use of prior elicitation widespread (yet)?}
\label{limitations}

Priors can have significant effect on the outcome of the whole modelling process and support is clearly needed for their specification \citep{Robert2007,ohagan:2019}, yet prior elicitation techniques are not routinely used within practical Bayesian workflows. The most natural explanation for this is that the current solutions are simply not sufficient for the needs of the people building statistical models and doing practical data analysis. We are not aware of structured literature looking into these aspects systematically, and hence we provide here our evaluation of the main reasons why prior elicitation has not yet entered daily use in the statistical modelling community. The goal here is to provide a high-level overview of the main issues we have identified based on both the scientific literature and our experiences while interacting with the modelling community, in particular the user bases of \texttt{Stan} \citep{stan}, \texttt{brms} \citep{brms1}, \texttt{PyMC} \citep{pymc} and \texttt{Bambi} \citep{bambi}. Not all claims of this subsection are supported by direct scientific evidence.

As briefly mentioned in the Introduction, we believe the reasons for limited use of prior elicitation are multifaceted and highly interconnected. We believe the three primary reasons, all of approximately equal importance, are: 
\begin{itemize}
    \item {\bf Technical:} We do not know how to design accurate, computationally efficient, and general methods for eliciting priors for arbitrary models.
    \item {\bf Practical:} We lack good tools for elicitation that would integrate seamlessly to the modelling workflow, and the cost of evaluating elicitation methods is high.
    \item {\bf Societal:} We lack convincing examples of prior elicitation success stories, needed for attracting more researchers and resources.
\end{itemize}

By the \emph{technical} dimension we refer to the quality and applicability of the prior elicitation methods and interfaces, for instance in terms of what kinds of models and priors are supported, and how accurate and efficient the algorithms are. An ideal solution would work in general cases, provide an easy interface for the expert to provide information, accurately reproduce the true knowledge of an expert, and be computationally efficient and reliable to be incorporated into the modelling workflow. In Section~\ref{summary_literature_review} we will summarize the current literature and discuss the limitations of the current technical solutions, effectively concluding that we do not yet have prior elicitation techniques that would reach a sufficient level of technical quality in general cases.

By the \emph{practical} dimension we refer to concrete tools ready to be used by practitioners.  On a rough level, a prior elicitation method consists of some interface for interacting with the expert and the computational algorithm for forming the prior. Often the interfaces proposed for the task have been fairly general, but the majority of the research on the computational algorithms has been dedicated to methods that are only applicable for specific models or forms of priors. Their practical value remains limited. 
Even though some examples of model-agnostic elicitation methods exist and some initiatives have been developed for extended periods of time, we are still nowhere near a point where prior elicitation tools would routinely be used as a part of the modelling process. Besides the technical reasons mentioned above, one major reason is that the tools have not been integrated as parts of the broadly used modelling ecosystems, but rather as isolated tools with their own interface conventions, modelling languages, and internal data formats. To put it briefly, a person building a model e.g. in \texttt{Stan} cannot launch an elicitation interface to elicit priors for their specific model, and in the extreme case there might not even exist any tools applicable to their model. In Section~\ref{software}, we will outline directions for overcoming this practical challenge.

Another practical issue concerns evaluation of prior elicitation methods. Even though the basis of evaluating the elicitation methodologies is well established (see Section~\ref{evaluation_elicitation}), the practical value of prior elicitation is extremely difficult and costly to evaluate. Already isolated studies demonstrating e.g. improved task completion time, compared to manual prior specification, for some prototypical model require careful empirical experimentation with human users. While this is a common practice in human computer interaction research, for statisticians it requires quite notable additional effort and expertise. More importantly, for the real cases of interest the evaluation setup is unusually complex because the modelling process itself is a highly complex iterative process that requires statistical expertise and takes a long time, possibly weeks or months. Any empirical evaluation of the value of prior elicitation requires enrolling high-level experts who are tasked to carry out complex operations with systems that are unfamiliar to them, and possible significant individual differences in the way models are built necessitate large user bases for conclusive evidence. This can only be done once the practical software is sufficiently mature, and even then is both difficult and expensive. The problem is naturally not unique to prior elicitation, but instead resembles e.g. the cost of evaluating the effect of new medical practices that require medical professionals testing new procedures that may also result in worse treatments, or evaluation of new educational policies and practices. However, justifying the cost is often easier for these tasks that are considered critically important for the society.

Following the above discussion on cost of evaluation, we believe that there is a significant \emph{societal} argument explaining the limited use of prior elicitation. As detailed in this article, the task is challenging and consequently requires significant resources spanning several scientific fields, combining fundamental statistical and algorithmic research with cognitive science and human-computer interaction for forming the solid basis with high-quality software integration and costly evaluation. This requires significant resources, yet the current research is driven solely by academia and the field has remained somewhat small. 

To some extent this can be attributed to the long history of avoiding strong subjective priors in quest for objective scientific knowledge or fair and transparent decision-making. Audiences struggling to accept subjective priors in the first place are best convinced by maximally clear examples that leave no room for additional layers of complexity, such as prior elicitation procedures. Follow-up research encouraged by these examples is likely to follow similar practices even when they could benefit from improved processes for prior specification. We hence argue that lack of broad interest more specifically on prior elicitation is largely because the value of prior elicitation has not been concretely demonstrated in breakthrough applications of societal importance. Without such demonstrations, the level of interest for these tools will remain low outside the statistics research community. However, already isolated examples of significant scientific or economical breakthroughs building on explicit use of prior elicitation could lead to increase in both research funding (e.g. in the form of public-private partnerships for applying the technology) and in particular in interest for open source software development. To some extent these efforts are shared with the general task of convincing researchers and decision-makers that use of subjective priors is scientifically valid and valuable, but additional effort is needed in demonstrating the value of prior elicitation, in the form of examples where it results in improved models or offers a more cost-efficient, reliable and reproducible process.

This argumentation, unfortunately, is very circular in nature. To boost interest in developing better prior elicitation methods, we would need a high-profile demonstration of their value, but establishing that demonstration would require access to high-quality solutions that integrate well with the modelling tools. However, it is important to realize that the demonstrator can likely be done well before having a robust general-purpose solution. Instead, it is sufficient to have proper software and interface integration of prior elicitation with one modelling ecosystem that is already used for addressing societally important modelling questions, combined with elicitation algorithms that work for the specific types of models needed and can later be extended for even more general models without changing the interfaces. For instance, Bayesian models developed within the \texttt{Stan} ecosystem played a significant role in modelling the effect of various interventions had on the spread of COVID-19 \citep{flaxman:2020}, and demonstrating the value of prior elicitation in such a context would likely have been sufficient for raising the awareness of this research direction.

\subsection{Prior elicitation hypercube}\label{prior_hypercube}

\begin{figure}
	\begin{center}
		\includegraphics[scale=0.44]{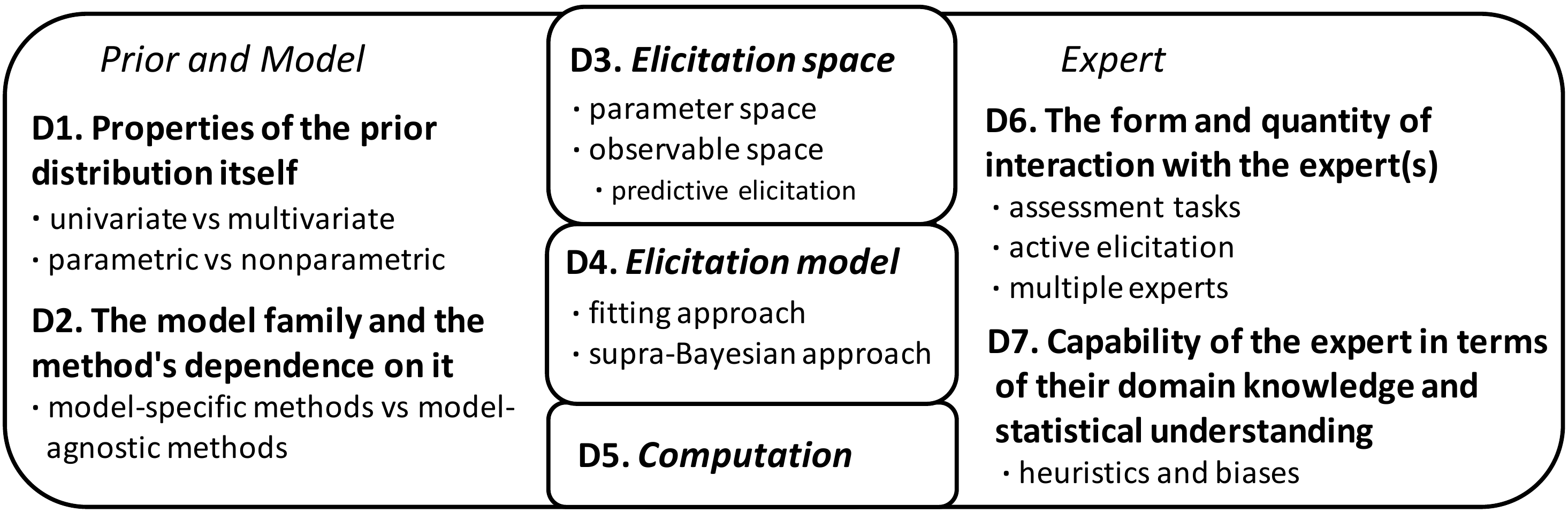}
	\end{center}
	\caption{\textbf{Prior Elicitation Hypercube}. The seven dimensions (D1-D7) of the hypercube.}\label{hypercube}
\end{figure}

The interdisciplinary nature of the prior elicitation problem, and therefore scattered coverage of the topic, makes it difficult to obtain an overall perspective to the current state of research. To provide a frame of reference, we identify seven key dimensions that characterize the prior elicitation problem. Together the dimensions form a \emph{prior elicitation hypercube}, depicted in Figure~\ref{hypercube}, that both helps discuss the current literature in a more structured manner and enables identifying understudied directions.
The first two dimensions (\hyperref[D1]{D1}-\hyperref[D2]{D2}) cover the Bayesian model itself (prior and likelihood). Dimensions \hyperref[D3]{D3}-\hyperref[D5]{D5} specify key proprieties of an elicitation algorithm, such as in which space the elicitation is conducted (\hyperref[D3]{D3}), how the expert's input is modelled (\hyperref[D4]{D4}), and how to deal with the computational issues (\hyperref[D5]{D5}). The last dimensions \hyperref[D6]{D6}-\hyperref[D7]{D7} cover what is assumed about the expert(s) and the interaction with them. The current prior elicitation literature is reviewed and categorized in terms of these dimensions in the Supplement. For convenience, the section numbers of the supplementary material are preceded by S, so that e.g. Section~S1 refers to the first section of the Supplement.

\textbf{D1: Properties of the prior distribution itself}\label{D1}. Two properties of the prior distribution have attained considerable attention in the literature: dimensionality and parametric vs nonparametric nature. Dimensionality is about the number of parameters: is the prior univariate or multivariate? Eliciting multivariate (joint) distributions is a more complex task than eliciting univariate (marginal) distributions \citep{ohagan:2006}. It is not enough to elicit the prior in a parameter-by-parameter manner because it is the joint behaviour that affects inferences, and hence it is the joint distribution that must be considered \citep{Gelman2017}. Maybe because of the challenge of eliciting multivariate priors, univariate elicitation has been studied more, even though most models have more than one parameter and hence multiparameter prior elicitation is really needed \citep[][Section 7.3]{Gelman2020}. 

The second property is about whether the prior comes from some parametric family or is nonparametric. The main strand of the prior elicitation literature is about the elicitation of parametric prior distributions, in which case the follow-up question is to which parametric family the prior belongs. The family is determined by the choice of the analyst rather than as a result of the elicitation, although it is often chosen so that it does not conflict with the elicitation data. The choice of the family is closely connected to the likelihood/model (see Section S4), since the natural conjugate family is often considered. On the other end, there is an important line of research on nonparametric prior elicitation that has been built upon Gaussian processes \citep{ohagan:2007}.

\textbf{D2: The model family and the method's dependence on it}\label{D2}. The underlying probabilistic model and the data analysis task in which it is applied, significantly impact the choice of the prior elicitation method. A bulk of the prior elicitation research addresses the elicitation of parameters of some specific model or model class. We call these types of methods \textit{model-specific}, and they are reviewed in Section S4. In contrast, in our literature review we found a relatively small number of \textit{model-agnostic} prior elicitation methods, but to name some, we refer the reader to \citet{gelfandetal:1995,ohagan:2007,hartmann:2020}. To promote adoption of prior elicitation in applications, it is highly desirable that a prior elicitation method is not model-specific, or at least is applicable to a wide range of models, and we strongly encourage research in this direction as acknowledged earlier by \citet{kadane:1998}. Furthermore, the underlying data analysis task may indicate which parameters are of interest, and thus need to be elicited (in the context of a chosen Bayesian model), and which may be less relevant (\citealt{clemen:2001}, p.292; \citealt{stefan:2020}).

\textbf{D3: Elicitation space}\label{D3}. Prior elicitation is about eliciting expert knowledge to form a prior distribution for model parameters. Hence, it is not surprising that the majority of prior elicitation research is focused on querying values of parameters, or quantities directly related to parameters, from the experts. In this case, we say that the underlying elicitation space is the \textit{parameter space}. This implies that the expert has to have at least some intuition about the meaning of the parameters (interpretability) and about their natural scales. However, this cannot be assumed in all cases. The elicitation of parameters of Bayesian neural networks serves as an extreme example. Models of this type can have thousands of parameters without any interpretation attached to them.  In many cases it may be more beneficial to query the expert about something else, such as model observables. In this case we say that the underlying elicitation space is the \textit{observable space}. The \textit{model observables} are variables (e.g.\ model outcomes) that can be observed and directly measured, in contrast to \textit{latent variables} (e.g.\ model parameters) that only exist within the context of the model and are not directly observed. \citet{kadane:1998} made a similar dichotomy where they called elicitation in the parameter space \textit{structural elicitation}, and elicitation where the expert makes judgments about ``the dependent variable given various values of the predictor variables'', \textit{predictive elicitation}. Predictive elicitation is a type of elicitation in the observable space. In general, elicitation in observable space does not require the model to have both dependent and independent variables \citep[e.g.][]{coolen:1992,hughes:2002,gaoini2009bayesian}, the existence of a ``regression likelihood'' \citep[][p.5]{kadane:1998}, nor the prior predictive distribution (S2.1). For instance, an `elicitation likelihood' can be used for connecting the expert's knowledge on observables to the parameters of interest (see Section~\ref{Bayesian_treatment_of_expert}).

\textbf{D4: Elicitation model}\label{D4}. There are fundamental differences between elicitation methods in terms of how the information provided by the expert is interpreted. Since early prior elicitation research \citep{winkler:1967,bunn:1978}, the dominant approach has been ``fitting a
simple and convenient distribution to match the elicited summaries'' \citep{ohagan:2006}. This `\textit{fitting approach}' does not assume any specific mechanism on how the expert data are generated, and for instance, inconsistencies in the data are reconciled by least-square minimization. \textit{Overfitting} in elicitation means eliciting more summaries than needed to fit a parametric distribution \citep{ohagan:2006,hosack:2017}, in which case inconsistencies may appear. Overfitting itself is desirable because it allows for imprecision in the elicited summaries, and the fitted compromise prior may be expected in practice to yield a more faithful representation of the expert’s knowledge \citep{ohagan:2006}.

There is an alternative to the fitting approach, and how inconsistencies are dealt with. The elicitation of an expert’s knowledge can be treated as any other Bayesian inference problem where the analyst’s posterior belief about the expert’s knowledge is updated in the light of received expert data \citep{lindley:1979,gelfandetal:1995,ohagan:2004,gosling:2005,ohagan:2007,daneshkhah:2006,gosling:2007,oakley:2010b,moala:2010,micallef:2017,hartmann:2020}. The analyst has his own prior over the expert belief, and there is an \textit{elicitation likelihood} that allows the analyst’s posterior, which would be the elicited expert's prior, to be inferred from the elicitation data. This standpoint is similar to supra-Bayesian pooling found in the literature of aggregating knowledge of multiple experts (Section S6). We follow the latter terminology, even if there is only a single expert to be elicited, and say that such an elicitation method follows the \textit{supra-Bayesian approach}. In this approach, inconsistencies in the elicited data are accounted for by a noise mechanism built into the elicitation likelihood. 

\textbf{D5: Computation}\label{D5}. Computation is needed in many parts of an elicitation algorithm, such as in constructing prior from the elicited data and in active elicitation (Section S5), and the computational aspects need to be accounted for in practical tools. One-shot (\hyperref[D6]{D6}) elicitation methods that follow the fitting approach (\hyperref[D4]{D4}) and solely operate in the parameter space are often computationally efficient and can easily be incorporated into a practical workflow. In contrast, iterative (\hyperref[D6]{D6}) predictive (\hyperref[D3]{D3}) elicitation methods that operate in both spaces and require repeated computation of the prior predictive distribution require considerably more attention in terms of computational efficiency, both because of increased computational cost and the need for fast response time for convenient user experience.

\textbf{D6: The form and quantity of interaction with the expert(s)}\label{D6}. The sixth dimension is about the interaction between the expert(s) and the analyst. On the one hand, the form of \textit{assessment tasks} that an expert performs (and similar aspects relating to the interaction modality with a single expert) is important. On the other hand, if there is more than one expert, the format in which the experts interact is also important. For instance, the behavioural aggregation method used in the SHELF protocol \citep{oakleyshelf:2019} encourages the experts to discuss their opinions, and to settle upon group consensus judgments, to which a single prior distribution is fitted \citep{ohagan:2019}. Eliciting the knowledge of a group of experts, and how to combine the elicited information into a single \textit{aggregate distribution}, is a well established topic.

Concerning a single expert, there are choices to be made about the interaction modality of the elicitation. The expert can be either queried in a one-shot manner (\textit{one-shot elicitation}), or iteratively where the expert's input affects what is queried next (\textit{iterative elicitation}). For instance, a prior elicitation algorithm that exploits active elicitation (Section S5) is iterative. We distinguish iterative elicitation from \textit{interactive elicitation} that entails interaction with the elicitation system \citep{kadane:1980}, such as the system updating a visualization of a prior distribution based on a slider position controlled by the expert \citep{jones2014prior}. It is not obvious at all in which form the information should be elicited from the expert. Several things need to be taken simultaneously into account, such as what assessment tasks are informative, computationally feasible, and, most importantly, encourage “a thoughtful, auditable and relevant answer that is not affected or biased in some way by the giver’s psychology” \citep{hanea:2021}. Thus, the design of the assessment tasks is key, as mathematically equivalent assessment tasks are not necessarily psychologically equivalent \citep{ohagan:2006}. For instance, the impact of the \textit{visualization} of assessment tasks on elicitation has been studied by \citet{hullman2018graphics},  \citet{kim2019bayesian,kim2020bayesian}, and \citet{sarma:2020}. Research on different assessment tasks is reviewed in Section S1. Since the assessment task should also consider psychological and cognitive aspects of a person being elicited \citep{ohagan:2019}, this topic is also related to the next dimension.

\textbf{D7: Capability of the expert in terms of their domain knowledge and statistical understanding}\label{D7}. 

Perhaps the most challenging and researched issue in prior elicitation is that most people are unable to express their prior knowledge in terms of probabilities, although they will do so if asked, but their answers may be based on very superficial thinking. \citep{hanea:2021,kahneman2011thinking}. If the expert has no solid statistical training, she may not be able to provide reliable probabilistic assessments. In that case, we can resort to assessment tasks that do not require probabilistic input, such as querying likely hypothetical samples \citep{casement:2018}. If the expert has only vague domain knowledge, the elicitation algorithm should validate the provided information, for instance, by using `seed variables' as in Cooke's method \citep{cooke:91}. Even if the expert has both excellent statistical and domain knowledge, she may be inclined to commit to popular cognitive biases and to use cognitive shortcuts (heuristics) in her reasoning, as well documented by \citet{tversky:1974}. This line of research is known as heuristics and biases in prior elicitation, and it is intrinsically connected to psychology \citep{hogharth:1975}. We provide only entry-points to this broad researched field in Section S7.

\subsection{Overview on past literature}\label{summary_literature_review}

We reviewed the current main lines of research in prior elicitation through the lens of the prior elicitation hypercube (Section~\ref{prior_hypercube}). The literature review can be found in the Supplement, with sections referenced to using S1, S2 and so on, and a summary of the main findings is presented here.

We observed that there are regions in the prior elicitation hypercube that are well understood. Elicitation of a univariate parametric prior is an extensively studied topic. Certain descriptive elements of the prior distribution, known as \textit{summaries}, such as quantiles, are typically queried from the expert. These univariate elicitation methods commonly differ in the type of elicited summary, the order the summaries are elicited in, and the framing of the corresponding assessment tasks (visual, gamble, etc.). The leading principle in thinking of the aforementioned aspects and designing elicitation methods in general, has been to minimize cognitive biases and so-called heuristics (Section S7) which expert probabilistic judgments may be subject to \citep{ohagan:2019}. There are widely accepted protocols on how to deal with these biases, and how to conduct elicitation with single (Section S1) and multiple experts (Section S6). However, not all methods take them properly into account.

The research on elicitation in the space of observables is abundant (Section S2), but with a serious limitation. Namely, almost all the research is model-specific. Some prior and model families have been studied extensively (Section S4), with significant attention e.g. on elicitation of priors for generalized linear models. From the perspective of priors, there have been several works on the specific cases of Beta and Dirichlet distributions (Section S1.3). When these priors are considered together with their conjugate likelihood, which allows for a complete sampling model, the assessment tasks are often in the space of observables. If this is not the case, then the assessment tasks are in the space of parameters.

There are also distinct research lines on scoring rules (Section S1.4), nonparametric elicitation (Section S3), and active elicitation (Section S5). \textit{Active elicitation} research refers to several articles on how active learning \citep{cohn:1994} can be applied in prior elicitation to help make most out of the limited elicitation budget due to costly human effort. \textit{Nonparametric prior elicitation} research is mostly built upon a supra-Bayesian elicitation framework, where the expert’s subjective density is assumed to follow a Gaussian process \citep{ohagan:2007}. \textit{Scoring rules} are a class of devices for eliciting and evaluating probabilities \citep{murphy:1970}. They encourage the expert to make careful assessments. 

Despite the fact that multivariate prior elicitation has been studied from many perspectives, many of these methods are difficult to apply beyond text book examples. In particular, the methods do not scale well to high-dimensional parameter spaces. Copula-based elicitation requires assessment of parameter dependencies, which is cognitively challenging \citep[][Sec. 2.3]{garthwaite:2005} and scales poorly (e.g.\ Gaussian copula requires specification of a covariance matrix, \citealp{clemen:1999}, with $dim(\boldsymbol{\theta})(dim(\boldsymbol{\theta})+1)/2$ elements). Nonparametric Gaussian process elicitation that in principle could work with higher dimensions has been empirically demonstrated only for two parameters \citep{moala:2010}. Predictive elicitation with generalized linear models \citep{kadane:1980,bedrick:1996} does not help either. Although the original method by \citet{kadane:1980} can handle linear regression on at least four covariates, scaling the method to hundreds of covariates is out of question due to the increasing number of elicitation queries. Furthermore, the independence assumption of covariates is troublesome in some predictive methods \citep{garthwaite:1988,bedrick:1996}. The fundamental challenge for these methods, and for multivariate methods in general, is how to find assessment tasks that are both feasible for the expert and informative enough to identify the complex joint prior distribution of parameters. Moreover, an inference algorithm is needed that can form a prior from the elicited data.

\section{Where Should We Be Going?}
\label{sec:future}

We have discussed some limitations of the current prior elicitation research (Sections \ref{limitations} and \ref{summary_literature_review}). In this section, we discuss possible solutions. We propose five promising avenues (Sections \ref{Bayesian_treatment_of_expert}-\ref{software}) to help in solving the technical, practical, and societal challenges described in Section~\ref{limitations}; we believe research on these avenues will increase the adoption of prior elicitation techniques.

\textit{Technical solutions}: We believe that an elicitation method should support elicitation both in the parameter and observable space, should be model-agnostic, and should be sample-efficient since human effort is costly. In Section~\ref{Bayesian_treatment_of_expert}, we propose an approach for prior elicitation that takes these objectives into account. We also believe that elicitation is easier when the prior is globally joint. These globally joint priors are discussed in Section~\ref{better-priors}, but essentially, they let elicitation be reduced to just a few interpretable hyperparameters.

\textit{Practical solutions}: To help make model building easier, faster and better in reflecting expert knowledge, we need to integrate prior elicitation into the Bayesian workflow (Section~\ref{bayesian_modelling_workflow}). And this requires software able to inter-operate with already existing tools for Bayesian modelling, including probabilistic programming languages (Section~\ref{software}). The software needs to support model-agnostic elicitation, otherwise there will be problems with integration into the Bayesian workflow, because a change in the model specification could preclude prior elicitation.

\textit{Societal solutions}: We emphasize the need for considerably extended user evaluation, required for verifying that the methods have practical value (Section~\ref{evaluation_elicitation}), and the need of case studies showing the advantages that a careful prior elicitation process can bring to the modelling process.

\subsection{Bayesian treatment of the expert in prior elicitation}\label{Bayesian_treatment_of_expert}

In this section, we propose a unified approach to prior elicitation that brings together several elicitation methods. The approach allows the expert to provide her response in both the parameter and observable space (\hyperref[D3]{D3}), and supports sample-efficient elicitation (\hyperref[D6]{D6}) by treating the expert in a Bayesian fashion.

In the supra-Bayesian approach, elicitation of an expert’s knowledge is treated as any other Bayesian inference problem where the analyst’s posterior belief about the expert’s knowledge is updated in the light of received expert data (see the discussion in \hyperref[D4]{D4}). We propose viewing the prior elicitation event itself as an interplay of the expert and the analyst with the following characteristics:
\begin{description}
\item[Analyst] Poses queries to the expert and gathers the expert's input into a dataset $\mathcal{D}$. The analyst's goal is to infer the expert's distribution of the parameters $\theta$, conditional on the expert's input data, $p(\theta | \mathcal{D})$.
\item[Expert] Based on her domain expertise, the expert answers to the analyst's queries. The expert's input is modelled through the \textit{user model} $p(z|q)$ that is the conditional probability of the expert's input $z$ given the analyst's query $q$. That is, $\mathcal{D}$ consists of $N$ samples $(z_i,q_i)_{i=1}^{N}$, and all the $q_i$ are treated as fixed.
\end{description}
Expert data can be provided in multiple elicitation spaces, all of which can be combined to derive a single prior within the user model. For example, we can elicit expert data in both the observable space (data $\mathcal{D}_{\mathcal{Y}}$) and in the parameter space (data $\mathcal{D}_{\Theta}$). The analyst's goal is then to infer the distribution of the parameters conditional on the expert's input data, that is $p(\theta | \mathcal{D}_{\mathcal{Y}}, \mathcal{D}_{\Theta})$. We assume that the analyst updates his knowledge according to Bayes’ rule. Hence, he treats elicitation as a posterior inference problem,
\begin{equation}\label{Bayes_eq}
p(\theta | \mathcal{D}_{\mathcal{Y}}, \mathcal{D}_{\Theta}) = \frac{p(\mathcal{D}_{\mathcal{Y}}| \theta)p( \mathcal{D}_{\Theta} | \theta)p(\theta)}{p(\mathcal{D}_{\mathcal{Y}}, \mathcal{D}_{\Theta})},
\end{equation}
given the elicitation likelihoods $p(\mathcal{D}_{\mathcal{Y}}| \theta)$ and $p( \mathcal{D}_{\Theta} | \theta)$, and the analyst's prior belief on the expert's knowledge $p(\theta)$. 
In Equation \ref{Bayes_eq}, we have assumed $\mathcal{D}_{\mathcal{Y}}$ and $\mathcal{D}_{\Theta}$ to be conditionally independent given $\theta$. The likelihoods $p(\mathcal{D}_{\mathcal{Y}}| \theta)$ and $p( \mathcal{D}_{\Theta} | \theta)$ account for the uncertainty inherent to the elicitation process due to the mechanism how the expert quantifies her knowledge on $\theta$. Hence, the conditional independence assumption essentially states that: given that there exists a fixed parameter vector $\theta$ that the expert thinks to be `true', the mechanism how the expert reveals her knowledge on $\theta$ is independent between the two elicitation spaces.

The analyst's prior $p(\theta)$ can be taken to be one of the `objective' priors mentioned in Section \ref{What_is_prior_elicitation}. Besides the prior, the framework only requires specifying $p(z|q,\theta)$ which describes, at individual query $q$ level, how the expert would respond if she thinks that $\theta$ is true. This $p(z|q,\theta)$ is also the likelihood for a single data-point $(z,q)$, since $q$ is treated as fixed without a probability distribution assigned to it. The user model can be obtained by marginalization, $p(z|q) = \int p(z|q,\theta)p(\theta)d\theta$.

The proposed approach can be readily extended to support both sample-efficient elicitation (via active elicitation) and AI-assisted elicitation. 

\textit{Active elicitation}. How to make the most out of the limited budget of $N$ expert's inputs? In other words, what is an optimal strategy to select a sequence of queries $(q_i)_{i=1}^{N}$? This is where the user model comes to play. When the analyst poses a query $q$, he anticipates that the expert's input $z$ is distributed according to $p(z|q)$. The analyst applies the user model to choose the most informative queries. For instance, if the analyst wants to maximize the expected information gain of $p(\theta | \mathcal{D})$ with respect to a new query $q$, then the user model is needed for anticipating the corresponding yet unseen response $z$, which involves taking expectation over $p(z|q)$.

\textit{AI-assisted elicitation}. One important thing to note is that the analyst (or here facilitator) need not manually select the next queries, but the whole elicitation process can be supervised by an `artificial facilitator' -- an \textit{AI-assistant}. For instance, the AI-assistant can be as simple as consisting only of a user model combined with an active learning criterion for selecting next queries. However, in principle, it is possible to extend the functionalities and capabilities of the AI-assistant to take into account, for instance, the expert's biases and incapabilities of providing informative input for some queries.

Through the following examples, we illustrate how the proposed approach brings together prior elicitation methods found in the literature:

\begin{itemize}
    \item \textit{Quantiles with mixture beta assumption} \citep{gelfandetal:1995}. $\mathcal{D}$ is a set of quantiles of the prior distribution of parameters. The elicitation space is the parameter space, $\mathcal{D}=\mathcal{D}_{\Theta}$. The likelihood $p( \mathcal{D}_{\Theta} | \theta)$ equals Eq.\ (4) in \citet{gelfandetal:1995}, and it is derived from a few assumptions, one being that the expert's input is a transformation of a mixture of beta-distributed random variables. The authors proposed using Markov chain Monte Carlo for sampling from the posterior $p(\theta | \mathcal{D}_{\Theta})$.
    \item \textit{Judgements about plausible outcomes} \citep{hartmann:2020}. $\mathcal{D}$ is a set of prior predictive probabilities where the expert provides $\textrm{P}(A_i|\lambda)$ for all $i=1,...,n$, given a partition $\mathbf{A} = \{A_1,...,A_n\}$ of the observable space and hyperparameter vector $\lambda$ of a parametric prior $p(\theta|\lambda)$. The elicitation space is the observable space, $\mathcal{D}=\mathcal{D}_{\mathcal{Y}}$. \citet{hartmann:2020} assumed a Dirichlet likelihood and used maximum likelihood estimation to estimate $\lambda$.
    \item \textit{Judgements about parameter values and relevance, using active elicitation} \citep{daee:2017}. The assumed model-specific setup considers a linear regression with a sparsity-inducing spike-and-slab prior \citep{george1993variable} on the regression coefficients. $\mathcal{D}$ is a set of judgements on regression coefficient values and relevance. The elicitation space is the parameter space, $\mathcal{D}=\mathcal{D}_{\Theta}$. The elicitation likelihood $p( \mathcal{D}_{\Theta} | \theta)$ and the analyst's prior $p(\theta)$ can be written as a product of Normal and Bernoulli distributions \citep[Appendix A]{daee:2017}. 
    
    The active elicitation approach in the paper mixes the regression data and the elicitation data. The proposed active elicitation criterion maximizes the information gain between the posterior predictive distribution and the posterior predictive distribution with a new expert’s data point $(z,q)$. The posterior predictive distribution is conditional to both the observational and elicitation data.
\end{itemize}

\subsection{Bayesian modelling workflow}\label{bayesian_modelling_workflow}

Having to choose a prior distribution can be portrayed both as a burden and a blessing. We choose to affirm that it is a necessity. If you are not choosing your priors yourself, then someone else is inevitably doing it for you, and the automatic assignment of flat priors is not a good idea \citep{carlin:2000,VANDONGEN:2006,gelman2006prior,BDA:2013,Gelman2017,smid2020dangers, martin2021}. Under some scenarios, we can rely on default priors and default models. For instance, we may simply need to use a given model for routine inference over new datasets. However, having the flexibility to alter model assumptions could be advantageous, and priors are just one form of assumptions. Thus, adopting a Bayesian workflow for prior elicitation should help to reduce the burden and increase the blessing.

We need a Bayesian workflow, rather than mere Bayesian inference, for several reasons \citep{Gelman2020}: Bayesian modelling can be challenging and generally requires exploration and iteration over alternative models, including different priors, in order to achieve inference that we can trust. Even more, for complex problems we typically do not know ahead of time what model(s), that is, the combination of prior and likelihood, we want to fit and even if so, we would still want to understand the fitted model(s) and its relation to the data. Such understanding can often best be achieved by comparing inferences from a series of related models and evaluating when and how conclusions are similar or not. 

One common practical approach to modelling starts with a template model \citep[see discussion by][]{Gelman2020} with default priors. A need for a more carefully designed prior may be revealed only after careful analysis of the first models, and it may be motivated by unrealistic results, computational problems, or the need for incorporating domain knowledge into a model. In other words, the choice of prior, as with other modelling decisions, is often informed by iterative model exploration. Prior elicitation is thus a central part of a Bayesian workflow, and is not restricted to happen only at the beginning of the workflow.

A useful workflow does not just follow all pre-described steps, but also omits them when they are unnecessary, in order to help allocate finite resources where they are most needed. For example, for simple parametric models and informative data, the likelihood can dominate the prior and the gain from prior elicitation could be negligible. Thus, in many cases it may be sensible to start with some common default priors or priors weakly informed by some summary statistics of the data (e.g. by centering and normalizing the covariate and target values in regression), and then assess the need for more careful prior elicitation using prior diagnostic tools \citep{kallioinen2021detecting}.

In that sense, knowing  when to perform prior elicitation is central to a prior elicitation workflow. A good general heuristic is ``in situations where prior information is appreciable, and the data are limited'' as \citet{ohagan:2006} have put it. Then, whether we should perform prior elicitation can be reformulated into: Is it worthwhile to spend resources to incorporate domain knowledge? Or more nuanced: How much information do we need to gather, and how accurate should that information be? In many instances, getting the order of magnitude right and/or obtaining a prior that works to remove nonsensical outcomes may be sufficient. Furthermore, the level of accuracy does not need to be the same for all the parameters in a model, as refining a few or even just one prior can translate into considerably better inference.

Informative priors are useful for inducing strong regularization, namely shrinkage priors such as horseshoe, regularized horseshoe, R2D2, spike-and-slab, and global-local-shrinkage priors. These are  applied, for example, in genetic association studies \citep{guan:2011} where there are a lot of covariates of which only very few are actually relevant and comparably small data sets, making inference without regularization very hard if not possible ($p \gg n$ problems, see \citet{peng:2013}). Outside such shrinkage priors and Bayesian trial design (since it is almost always a small-data scenario, see \citet{yuan2016bayesian}), there can be more nuanced scenarios where informative priors via prior elicitation are crucial. For example, there can be gaps in time-series data in which case the expert may provide structural information in a form of a prior distribution that helps to fill gaps in the posterior distribution, or the expert knowledge may help to extrapolate from one group in the data to another \citep[e.g. see][]{siivola:2021}.

In line with the current literature, we have so far discussed prior elicitation with regard to the choice of distributions and their parameters. This definition can be naturally extended to prior elicitation over models, which could provide a new sub-field for prior elicitation or a sister field of model elicitation. As evaluating over the entire range of conceivable models is unfeasible, answering questions such as: ``Is a linear model adequate?'', ``Do we need to extrapolate and perform predictions outside the observed domain?'', and similar ones would help us to narrow down options and save resources. Restricting the search to a few options early on will help, even if we later choose to expand the set of models.

Finally, a prior elicitation workflow should include one step to assess that the incorporated information is actually useful and an evaluation of the sensitivity of the results to the prior choice, including possible prior-data conflicts \citep{depaoli2020importance,Gelman2020,lopes2011confronting,al2017optimal,evans2006checking,reimherr2021prior,berger1990robust,berger1994overview,canavos1975bayesian,hill1994sensitivity,skene1986bayesian,jacobi2018automated,roos2015sensitivity,perez2006mcmc,giordano:2018,bornn2010efficient,ho2020global,kallioinen2021detecting}.

\subsection{Developing better priors} 
\label{better-priors}

One direction to improve prior elicitation is to develop priors for which elicitation is easier \textit{per se}.  In this context, `easier' can mean one of at least three perspectives: (a) easier to understand for experts (\hyperref[D7]{D7}), (b) computationally easier (\hyperref[D5]{D5}), and/or (c) leaving 
fewer degrees of freedom, that is, fewer  hyperparameters to elicit. Perspective (a) is especially relevant for direct elicitation in the parameter space, while perspective (b) is mostly relevant for indirect elicitation in the observable space due to computational requirements of the translation procedure to the parameter space ((\hyperref[D3]{D3}); see Section S2). Both of these perspectives tend to go hand in hand with the perspective (c) because fewer required choices often make the priors easier to understand for experts due to reduced cognitive load, and reduce computational requirements due to a  lower-dimensional target space of the translation. Accordingly, if we focus on (c), we can have the justified hope that other advantages will naturally follow in the process.

Reducing the number of hyperparameters comes with the initial (model-building) choice of what matters to be elicited and what is acceptable to just fix to a constant or forced to be of the same value (equality constraint). This line of reasoning leads to the notion of \emph{joint} hyperparameters where the individual priors all depend on a much smaller (or highly structured) set of hyperparameters, jointly shared across parameters. Any kind of \emph{hierarchical prior} follows this logic by design \citep{brms1}. For example, consider a simple hierarchical linear model across observations $i$ with intercepts $a_j$  varying across a total number of $J$ groups:
\begin{align*}
    y_i &\sim \text{normal}(\mu_i, \sigma) \\
    \mu_i &= a_{j[i]} \\
    a_j &\sim \text{normal}(a, \tau) \\
    a &\sim \text{normal}(\mu_a, \sigma_a) \\
    \tau &\sim \text{Gamma}(\alpha_\tau, \beta_\tau)
\end{align*}
Focusing on the priors for $a_j$, we have essentially reduced the problem of finding a total of $J$ priors, each with one or more hyperparameters, to just choosing four hyperparameters, namely the location $\mu_a$ and scale $\sigma_a$ of the normal prior on the joint mean $a$ as well as the shape $\alpha_\tau$ and rate $\beta_\tau$ of the Gamma prior on the joint standard deviation $\tau$.\footnote{In hierarchical models, it is common to also call $a$ and $\tau$ `hyperparameters' although they are not set by the analyst or expert but rather estimated from the data along with other model parameters. To avoid confusion, we continue to restrict the use of `hyperparameters' to parameters chosen in the elicitation process, which are thus fixed during model fitting.} 
However, such hierarchical priors are only \emph{locally joint} in the sense that they 
do not encompass all or even most parameters but only a subset. This becomes apparent if we extend the above model by additional additive terms, for example,
\begin{equation*}
    \mu_i = a_{j[i]} + b_i + c_i + d_i ,
\end{equation*}
with each term having their own mutually independent set of parameters and corresponding hyperparameters. 

It would be desirable to develop priors that are \emph{globally joint} in that they span most or even all parameters leaving just a few hyperparameters to choose. With the purpose of preventing overfitting and facilitating variable selection in high-dimensional linear regression models on comparably sparse data, several \emph{hierarchical shrinkage priors} have been developed that fulfil these properties \citep{bhattacharya:2015, piironen:2017, zhang:2020}. However, they do not yet generalize much beyond linear regression settings and their usefulness in the context of prior elicitation has not been studied so far. If we can extend these priors to more complicated models and find parameterizations with intuitive hyperparameters, such globally joint priors could prove extremely valuable in making prior elicitation more practical and widely applicable.

\subsection{Evaluating prior elicitation}\label{evaluation_elicitation}

When any new prior elicitation method is proposed, a natural question that arises is whether it works as desired. Similarly, when a variety of prior elicitation methods are available for a given context, the practitioner wonders which one is better. Such questions concern the evaluation of prior elicitation methods. There are multiple desiderata for prior elicitation. \citet{johnson:2010review}, for instance, categorize these into (i) validity -- whether the elicitation captures the true belief of the expert, (ii) reliability -- whether repeated elicitations reproduce the same priors, (iii) responsiveness -- whether the elicitation is sensitive to changes in beliefs, and (iv) feasibility, which refers to the costs or resources required for elicitation. Many of these desiderata may seem as being at odds with each other, but they are all relevant for the eventual goal of supporting the building of good models with available resources.

In an ideal scenario, any researcher or user of prior elicitation methods would easily be able to compare the pros and cons of existing off-the-shelf methods for her problem, or even test new ones in small-scale user studies. So far, there have been very few projects where multiple prior elicitation methods have been  empirically compared \citep{ winkler:1967,JOHNSON:2010,grigore2016comparison}, and these have been in very application specific contexts. There is a need for more general and  standard validation paradigms for prior elicitation, and the prior elicitation field has no equivalents to practices such as using benchmark datasets for comparing machine-learning algorithms, e.g. \citet{deng2009imagenet,lecun2010mnist}. We think this is a particularly challenging topic to work on because we also lack good metrics for evaluation. The simple metrics that have been widely used in this context may not be valid measures of the quantities we care about. For example, (i) an expert's subjective feedback about elicited priors may be subject to the kind of biases that also distort their priors, (ii) task completion time is considered to be a proxy for cognitive effort, but the elicitation may be finished inaccurately and in a hurried manner due to the cognitive strain it produces, and so on. Prior elicitation metrics can be potentially improved by incorporating research from areas such as Psychology and Human-Computer Interaction. Improved metrics, increased comparative work and the development of standardized validation paradigms or platforms would be essential as the prior-elicitation field makes more progress. In addition, many proposed evaluation metrics are model-specific, but we also need more general methods that can be used across the board in a model-agnostic manner.

Among the different criteria for prior elicitation, assessing faithfulness, accuracy, or validity may be the hardest. From this perspective, the aim of prior elicitation is to accurately capture subjective knowledge of experts/users. However, there are many sources of distortions in priors elicited by an expert including their cognitive biases while making judgments in uncertain settings, and measurement noise introduced by the prior elicitation method, for example, by eliciting probability distributions over discretized intervals  \citep{miller1983discrete, parmar1994chart, tan2003elicitation}, especially when there are a smaller number of intervals or bins. A promising empirical approach to evaluating faithfulness of prior elicitation would involve validating elicited priors against an expected ground truth. For instance, one could train participants on data produced by a specified model with specified priors, and see how well the true parameter priors are recovered by the elicitation methods. Such methods could be the basis for developing test-beds for prior elicitation evaluation. 

Model-specificity and training efforts in paradigms to evaluate faithfulness can also be bypassed by comparing elicited results against a `gold standard' model-agnostic method, which is known to have higher accuracy. While the nature of such baseline methods would be a topic of future research, there may be some viable candidates.  A very promising perspective in psychology treats human judgements as a result of sampling from their subjective probabilities. This viewpoint has been successfully applied in the Markov chain Monte Carlo with people (MCMCP) approach \citep{sanborn2008markov, sanborn2010uncovering} and its variants \citep{hsu2012identifying, leon2020uncovering, harrison2020gibbs} to elicit beliefs  about how stimuli from a multidimensional stimulus space (e.g. n-dimensional stick figures) maps onto a target category (e.g. `shape of a cat'). In MCMCP participants take the place of an MCMC acceptance function, and repeatedly accept or reject proposals regarding the category membership of the sampled stimuli. The adaptive nature of MCMC ensures that proposals are over time increasingly sampled from parts of the stimulus space representing the participants' subjective representation of the category. The participants' prior beliefs are then constructed as the stationary distribution of the Markov chain that their judgments eventually converge to. The performance of MCMCP and its variants, on natural categories as well as trained artificial categories make us believe that similar sampling-based methods may have promise in the prior elicitation field both, for obtaining faithful priors, and for acting as model-agnostic baseline methods in paradigms that assess faithfulness. 

When evaluating the accuracy of prior elicitation, we may also want to consider the effect of the elicited prior on the predictions or decisions made based on the model. In some scenarios, it is possible that even coarse elicitation processes can obtain practically useful information and further refinement of the elicitation may not bring additional benefits. Also, even if there is a significant bias in the elicited prior, that bias may have negligible effect on the end result. It can thus be useful to evaluate sensitivity and robustness of inference with respect to the elicited prior and its potential aspects that are difficult to elicit. For example, it is difficult for humans to estimate tiny probabilities, which is reflected in the difficulties of determining the tail shape of the elicited prior. A bias in the elicited prior and too thin tails can lead to strong prior sensitivity or prior-data conflict \citep{al2017optimal, evans2006checking, kallioinen2021detecting, paulTalk}. On the other hand,  thick tailed priors may lead to ignoring the otherwise correctly elicited prior information.

\subsection{Software for Prior Elicitation}
\label{software}

The absence of general software for prior elicitation that integrates well with existing probabilistic programming languages and tools is hindering the adoption of Bayesian methods outside our core community, and is thus eventually detrimental to their wider development.  As with other tools designed to help with the Bayesian workflow, a general design guideline is to avoid automated solutions that could result in the user not paying proper attention to their choices. Ideal software for prior elicitation should take into account the  strengths and weaknesses of both humans and computers. Numerical tasks that are computationally demanding, error-prone or even tedious should be automatized as much as possible, while allowing the user to retain control of important decisions and, ideally, the user should be helped to take informed decisions and avoid mistakes. For example, a prior elicitation tool should help users to incorporate domain knowledge while preventing them to become overconfident about their own opinions, and it should easily integrate with other tools to perform prior sensitive checks, for example.

In addition to these general guidelines, there are several desirable features that a software for prior elicitation could have, such as being open source and having a simple and intuitive interface suitable for non-specialists. At least one part of such interface should be visual to enable better input from humans and to perform validation of the proposed priors, and some level of interactive visualization capability would further help to obtain information from experts. Furthermore, switching between different types of visualization (kernel density estimates plots, quantile dotplots, histograms, etc.) would also be valuable as would be the possibility to add user-defined transformations before visualization. For example, \citet{sarma:2020} describe how different visualizations could lead to different strategies for prior elicitation, and that most participants in their study primarily used a combination of strategies for determining their choice of priors. In addition, research shows that even people with statistical training can have problems correctly interpreting probability densities (Section S1.1), and so alternative representations like quantile dotplots may be preferred \citep{kay2016ish}.

Prior elicitation software could be written agnostic of the underlying programming language, or at least interoperable with as many languages as possible, in order to avoid duplication of efforts. Building on top of already present open source libraries related to Bayesian workflow and uncertainty visualization like \texttt{ggdist} \citep{Kay2021}, \texttt{Bayesplot} \citep{Gabry2019, Gabry2021} and \texttt{ArviZ} \citep{kumar:2019} would help to achieve this goal. Moreover, working on top of such libraries could help to maintain modularity, which is especially desirable at the present state of development of the software for prior elicitation. Modularity would also help to reduce computational costs, if experimentation with visualizations and transformations can be made independent of the model. By specifying each task as distinctly as possible and dividing work, the community can generate and maintain software more easily, while at the same time encouraging research in prior elicitation on one or several dimensions of the research hypercube. 

Given that we still need more research to assert which elicitation space (\hyperref[D3]{D3}) is more appropriate for a given research problem, building software in a modular fashion should allow users to switch between the parameter and the observable space as needed. Similarly, the type of assessment task (\hyperref[D6]{D6}) should be something that can be chosen by the user (e.g.\ as in 
\texttt{SHELF} or 
\texttt{MATCH}). It is also important to develop software that supports model-agnostic prior elicitation (\hyperref[D2]{D2}), otherwise there will be problems with integration into the Bayesian workflow (Section~\ref{bayesian_modelling_workflow}) because a change in the model specification could preclude prior elicitation.

\section{Conclusion}

This paper covered the state of the prior elicitation today, focusing on discussing reasons for the somewhat limited impact the research has had on practice. We identified bottlenecks at different levels and argued that significant coordinated effort covering several scientific disciplines will be needed to transform the current state and make prior elicitation a routine part of the practical modelling workflow. In summary, we make the following concrete calls to arms:
\begin{enumerate}
    \item {\bf We need to focus on elicitation matters that answer to the needs of practical modelling workflow}. Compared to past research, the efforts should be re-directed more towards (a) elicitation methods that are agnostic of the model and prior, (b) elicitation strategies (e.g. active elicitation) that are efficient from the perspective of the modeller and compatible with iterative model-building, and (c) formulations that make elicitation of multivariate priors easier, for instance by designing hierarchical priors that are simpler to elicit.
    \item {\bf We need better open software} that integrates seamlessly into the current modelling workflow, and that is sufficiently modular so that new elicitation algorithms can be quickly taken into use and evaluated in concrete modelling cases. The elements not specific to elicitation algorithms (e.g. visualization of the priors, the language used for specifying the models and desired prior families) should be implemented using existing libraries whenever possible, and the tools should be open source.
    \item {\bf We need cost-efficient and well-targeted evaluation techniques} for supporting development of new methods and validating their relative quality and value in practical tasks. In ideal case, we would like to see a testbed for prior elicitation techniques that enable easy evaluation of alternative methods in varying situations with feasible experimentation cost, as well as practical ways of collecting information about efficiency of elicitation methods in real use cases.
    \item {\bf We need spearhead examples} that clearly demonstrate the value of prior elicitation in applications of societal interest to increase enthusiasm beyond the current niche. These examples need to be ones where use of subjective prior knowledge is useful without a doubt and additionally prior elicitation either improves the value of the model over carefully crafted priors or results in clear cost reductions or improved robustness via a more efficient process (e.g. for cases where the priors need to be specified repeatedly or for several parallel cases).
\end{enumerate}
For the first two we already outline concrete directions in Section~\ref{sec:future}. We hypothesize that addressing all four foci will transform the status of prior elicitation, by providing the required infrastructure, public interest and funding for speeding up future development.

\begin{supplement}
\stitle{Supplementary Material and Literature Review}
\sdescription{In this supplementary material, we present the current main lines of research in
prior elicitation through the lens of the prior elicitation hypercube (Section \ref{prior_hypercube}).}
\end{supplement}

\bibliographystyle{ba}
\bibliography{refs}

\begin{acks}[Acknowledgments]
This work was supported by the Academy of Finland (Flagship program: Finnish Center for Artificial Intelligence FCAI), by the Technology Industries of Finland Centennial Foundation, by the Jane and Aatos Erkko Foundation, European Research Council grant 742158 (SCARABEE, Scalable inference algorithms for Bayesian evolutionary epidemiology), and by the UKRI Turing AI World-Leading Researcher Fellowship, EP/W002973/1. Partially funded by Deutsche Forschungsgemeinschaft (DFG, German Research Foundation) under Germany’s Excellence Strategy - EXC 2075 – 390740016.
\end{acks}

\end{document}



\begin{frontmatter}
\title{Supplementary Material and Literature Review for\\  Prior knowledge elicitation:\\ The past, present, and future}
\runtitle{}

\begin{aug}
\author{\fnms{Petrus} \snm{Mikkola}\thanksref{addr1}\ead[label=e1]{petrus.mikkola@aalto.fi}},
\author{\fnms{Osvaldo A.} \snm{Martin}\thanksref{addr1,addr6}\ead[label=e2]{osvaldo.martin@aalto.fi}},
\author{\fnms{Suyog} \snm{Chandramouli}\thanksref{addr1,addr2}\ead[label=e3]{suyog.chandramouli@aalto.fi}},
\author{\fnms{Marcelo} \snm{Hartmann}\thanksref{addr2}\ead[label=e4]{marcelo.hartmann@helsinki.fi}},
\author{\fnms{Oriol} \snm{Abril Pla}\thanksref{addr2}\ead[label=e5]{oriol.abrilpla@helsinki.fi}},
\author{\fnms{Owen} \snm{Thomas}\thanksref{addr3}\ead[label=e6]{o.m.t.thomas@medisin.uio.no}},
\author{\fnms{Henri} \snm{Pesonen}\thanksref{addr3}\ead[label=e7]{h.e.pesonen@medisin.uio.no}},
\author{\fnms{Jukka} \snm{Corander}\thanksref{addr3,addr2,addr7}\ead[label=e8]{jukka.corander@medisin.uio.no}},
\author{\fnms{Aki} \snm{Vehtari}\thanksref{addr1}\ead[label=e9]{aki.vehtari@aalto.fi}},
\author{\fnms{Samuel} \snm{Kaski}\thanksref{addr1,addr5,equalcont}\ead[label=e10]{samuel.kaski@aalto.fi}},
\author{\fnms{Paul-Christian} \snm{Bürkner}\thanksref{addr4,equalcont}\ead[label=e11]{paul-christian.buerkner@simtech.uni-stuttgart.de}}
\and
\author{\fnms{Arto} \snm{Klami}\thanksref{addr2,equalcont,correspon}\ead[label=e12]{arto.klami@helsinki.fi}}


\address[addr1]{Helsinki Institute of Information Technology, Department of Computer Science (PM,OAM,AK,SK) \textcolor{white}{XX} and Department of Communications and Networking (SC), Aalto University, Finland}

\address[addr2]{Helsinki Institute of Information Technology, Department of Computer Science (MH,OA,AK,SC) \textcolor{white}{XX} and Department of Mathematics and Statistics (JC), University of Helsinki, Finland    
}

\address[addr3]{Institute of Basic Medical Sciences, University of Oslo (JC,HP) and\\  \textcolor{white}{XX} Akershus Universitetssykehus (OT), Norway
}

\address[addr4]{Cluster of Excellence SimTech, University of Stuttgart, Germany
}

\address[addr5]{Department of Computer Science, University of Manchester, UK}

\address[addr6]{Instituto de Matemática Aplicada San Luis, CONICET-UNSL, Argentina}

\address[addr7]{Parasites and Microbes, Wellcome Sanger Institute, UK}

\address[equalcont]{Equal contribution}
\address[correspon]{Corresponding author, \printead{e12}}

\end{aug}


\end{frontmatter}


\section*{Current Main Lines of Research}\label{literature_review}
Prior elicitation has a long history dating back to the 1960s when \citet{edwards:1963} introduced ``Problem of prior probabilities'' to the psychological community, followed by a prior elicitation article by \citet{winkler:1967}. The interdisciplinary topic spans several fields including statistics, psychology, economics, decision sciences, and more recently, machine learning. Similarly, there are numerous applications in many fields including clinical and pharmaceutical research \citep{best:2020,yuan2016bayesian,thall2003dose,alhussain:2020,ziyad:2020,montague:2019,grigore2013review,wolpert:1989,haakma:2011,tompsett:2020,legedza:2001}, management sciences \citep{galway:2007,kallen:2009}, environmental sciences \citep{wolfson:1996,coles:1996,hirsch:1998,choy:2009,hammond:2001,alawadhi:2006,leon2003elicitation}, social sciences \citep{gill:2005,LOMBARDI:2012,DELNEGRO:2008}, business \citep{crosby:1980,bajari:2003,Ming:2016} and physics \citep{craig:1998}. This complicates systematic literature review, as prior elicitation is addressed from different perspectives, in different contexts, and under different terminology. In decision sciences, elicitation research has been conducted using the phrases 
`expert judgement', `expert knowledge elicitation' and `probability encoding' \citep{spetzler:1975,dias:2018,bocker:2010,hanea:2021}. In statistics, the main branch of the prior elicitation research is presented in \citet{ohagan:2006}, \citet{garthwaite:2005}, and \citet{ohagan:2019}. For other relevant reviews, see \citet{chesley:1975}, \citet{wolfson1995elicitation}, \citet{jenkinson:2005}, \citet{european:2014}, \citet{oakleyshelf:2019}, and \citet{stefan:2020}, and in clinical research, see \citet{johnson:2010review} and \citet{azzolina2021prior}, and in applied machine learning, see \citet{kerrigan2021survey}.

In this supplementary material, we present the current main lines of research in prior elicitation through the lens of the prior elicitation hypercube (Main Section 2.3). The literature review was done in two stages. For forward reference searching, we used the Google Scholar search engine. The search strings used were ``\textsc{allintitle: prior elicitation}'', ``\textsc{allintitle: prior assessment}'', and ``\textsc{allintitle: (elicitation or eliciting) (distribution or distributions)}''. The searches were conducted on March 2021, produced in total 133 + 1000 + 182 = 1315 results. Our inclusion criteria were (i) published work, (ii) available, and (iii) pertinent to methodological prior elicitation. In exceptional cases, we included work that did not meet all the criteria (i)-(iii). The criterion (ii) means work which can be found in the public web. The two most common reasons of failing the criterion (iii) were that the work was an application of an existing prior elicitation method, and the work did not address the elicitation at all (e.g.\ papers about prior specification). In the backward reference searching stage, we identified relevant research lines and references on prior elicitation by starting search from the references in the papers found in the first stage. The majority of relevant references were found in the backward reference searching stage. Indeed, the used search strings turned out to reach only a small fraction of the overall prior elicitation research.

\section{Elicitation in parameter space}\label{parameter_space_sec}

This section discusses prior elicitation in the space of parameters (Hypercube D3). The goal is to form $p(\theta)$\footnote{Recall the notation in the main text Equation 2.1.} by eliciting information on $\theta$ from the expert. However, many of the  methods can be equivalently applied to eliciting information on $y$, if combined with a suitable algorithm to construct $p(\theta)$ based on the obtained information, discussed in Section~\ref{observables_space_sec}.

The first three subsections are organized according to the prior properties (Hypercube D1): univariate prior, multivariate prior, and some prior-specific elicitation methods. An introduction to elicitation in univariate parameter spaces can be found in \citet{oakley:2010}, and in multivariate parameter spaces in \citet{daneshkhah:2010}. The last subsection discusses a popular technique for evaluating the performance of a prior elicitation algorithm. 

Within the subsections themselves, we further organize the content according to the types of assessment tasks (Hypercube D6). The basic assessment task involves assessing quantiles in all three well-known elicitation protocols: Sheffield Elicitation Framework (SHELF) \citep{oakleyshelf:2019}, Cooke protocol  \citep{cooke:91}, and Delphi protocol  \citep{european:2014}. Specifically, SHELF incorporates many common assessment tasks such as (i) plausible limits, (ii) median, (iii) quartiles, (iv) tertiles, (v) roulette, (vi) quadrant probabilities, and (vii) conditional range. The tasks (ii)-(v) amount to asking for different types of quantiles that are particularly relevant when the prior is univariate, and they are discussed next. Subsequently, the tasks (vi)-(vii) that are relevant for a multivariate prior, are discussed.

\subsection{Univariate prior}\label{Quantiles,moments...}

Let us consider assigning a probability distribution for a scalar $\theta$. Representing one’s knowledge in probabilistic terms is known to be challenging \citep{tversky:1974,ohagan:2019}, let alone asking directly for the full density function $p(\theta)$ \citep{ohagan:1988,goldstein:2007}. For this reason, a major strand in the prior elicitation literature approaches this problem by asking certain descriptive elements of the distribution of $\theta$ from the expert. The descriptive elements, known as \textit{summaries}, commonly include quantiles such as quartiles and medians, moments such as means, and limit values such as lower and upper bounds of $\theta$. Prior elicitation reduces to eliciting a number of summaries of $p(\theta)$, and if well-chosen, the summaries should be enough to identify $p(\theta)$ \citep{ohagan:2006}. 

What summaries to elicit? There are plenty of psychological studies of people's ability to estimate statistical quantities such as measures of central tendency. For an overview of relevant research in this area, see \citet[][Section 2.2]{garthwaite:2005}, \citet[][Section 5.2]{ohagan:2006}, and \citet[][Section 6]{ohagan:2019}. The literature broadly agrees on a few aspects. Mode and median are preferred over mean, since there is strong evidence that if the population distribution is highly skewed, then people's assessments of the mean are biased toward the median \citep[e.g.][]{peterson:1964}. The higher-order moments (than expectation) should not be asked \citep{kadane:1998}. For example, \citet{garthwaite:2005} write about estimating the variance, ``people are poor both at interpreting the meaning of `variance' and at assigning numerical values to it'' \citep{peterson:1967}.

There are two popular elicitation methods that differ by the type of elicited summary. The first is the \textit{variable interval method} (or \textit{V-method}, \citealp{spetzler:1975}), which asks for quantiles from the expert. The second is the \textit{fixed interval method} (or \textit{P-method}, \citealp{spetzler:1975}), which asks for probabilities. For an overview, see \citet{oakley:2010}. In decision sciences, the methods are referred to as probability encoding methods \citep{haggerty:2011,spetzler:1975,hora:2007}. The methods were first introduced in the context of a temperature forecasting application \citep{murphy:1974}. Through experimentation involving 103 judges, \citet{abbas2008} showed ``slight but consistent superiority'' for the variable interval method along several dimensions such as accuracy and precision of the estimated fractiles.  It is also possible to jointly ask both quantiles and probability, a \textit{quantile-probability tuple}, in which case the method is referred to as \textit{PV-method} \citep{spetzler:1975}.

Variable interval method as presented by \citet{oakley:2010} and \citet{raiffa:1968} consists of eliciting the median, lower and upper quartiles, and possibly minimum and maximum of $\theta$. \citet{oakley:2010} divided the elicitation into seven steps. In the beginning, the analyst elicits the (i) median, (ii) lower quartile, and  (iii) upper quartile from the expert. It is possible to present assessment tasks in a form of gambles when the expert is not comfortable with directly providing quantiles. Then, the analyst asks (iv) the expert to reflect on her choices to check for consistency. Next, he (v) fits a parametric distribution (i.e.\ estimates its hyperparameters) by minimizing a sum of squared errors between the elicited quartiles and the quartiles of the cumulative distribution function of a parametric prior. The sixth step is `feedback stage' that involves (vi) presenting the fitted distribution back to the expert and allowing the expert to modify her original judgements. Similar quantile-based elicitation has been studied in \citet{garthwaite:2000} and \citet{dey:2007}.

Fixed interval method as presented by \citet{ohagan:1998} and \citet{oakley:2010} consists of eliciting lower and upper bounds, mode, and five probabilities of $\theta$. For instance, the probabilities $\textrm{P}(\theta_{min}<\theta< \theta_{mode})$ and $\textrm{P}(\theta_{min}<\theta< (\theta_{mode}+\theta_{min})/2)$ are elicited. Then, the analyst fits a parametric distribution by minimizing a sum of squared differences between elicited probabilities and the corresponding probabilities implied by the prior distribution. A default choice for prior distribution is a beta distribution with possibly scaled support $(\theta_{min},\theta_{max})$. Prior work on this method can be found in \citet{phillips:1993} and \citet{ohagan:1995}.

A closely related method is the \textit{histogram method}, which is popular in the expert knowledge elicitation applications \citep{smith1995subjective,forbes1994application,coolen1992bayes,onkal1996effects,roosen2001capturing,lahiri1988interest,pattillo1998investment,grisley1983farmers}. The method is simple, consisting of a few steps: The analyst asks the minimum and the maximum of $\theta$. Then, he splits the obtained range into a finite number of sub-intervals of equal length. Finally, he asks the expert to assign probabilities to each interval. The elicited histogram can be either used as a nonparametric prior (see Section~\ref{nonparametric_prior}) or a parametric prior distribution can be fitted by using some optimization procedure. An alternative, facilitated approach, is to ask the expert to construct the histogram by the aid of \textit{gambling devices} such as dice, coins, and roulette wheels. \citet{gore:1987} introduced the \textit{roulette method} in which the expert allocates chips to the bins. The probability of a parameter lying in a particular bin is interpreted as the proportion of chips allocated to that bin. \citet{JOHNSON:2010} verified `feasibility, validity, and reliability' of this method in an experiment with 12 subjects. The roulette method is supported in many elicitation tools, such as \texttt{SHELF} \citep{oakleyshelf:2019} and \texttt{MATCH} \citep{morris:2014}. 

\subsection{Multivariate prior}\label{multivariateprior}

Let us consider a vector of parameters $\boldsymbol{\theta} = (\theta_1,...,\theta_d)$ to which we want to assign a joint probability distribution. Given that there are available methods for eliciting univariate $\theta$, a popular strategy is to elicit $\boldsymbol{\theta}$ by multiple independent univariate elicitations. For this, $\boldsymbol{\theta}$ can be transformed into a new vector $\boldsymbol{\phi}$ in which the coordinates are independent. Alternatively, the analyst may elicit the conditional distributions of $\theta_i | \boldsymbol{\theta}_{-i}$ for all $i \in \{1,...,d\}$ with hypothetical values for the $\boldsymbol{\theta}_{-i}$. In some special cases it is possible to elicit marginals by using univariate elicitation methods, and then complete the multivariate prior by additional elicited information such as covariances and correlations between parameters. The drawback of this strategy is that correlations are difficult to assess \citep{jennings_amabile_ross_1982,wilson:1994}. For an overview of eliciting covariances and correlations, and in general the dependence, see \citet{werner:2017} and \citet{kurowicka:2006}.

In the context of multivariate elicitation, the process of transforming $\boldsymbol{\theta}$ into a new parameter vector with independent coordinates is called \textit{elaboration} \citep{oakleyshelf:2019}. The independence here refers to `subjective independence' meaning that the additional information about one parameter will not affect the expert’s opinion about the others. For instance, if $\theta_1$ and $\theta_2$ are treatment effects of two drugs, then the analyst may consider a transformation $\boldsymbol{\phi}=(\theta_1,\theta_2/\theta_1)$. The relative effect $\theta_2/\theta_1$ is likely to be subjectively independent of $\theta_1$, since the expert can consider separately the magnitude of $\theta_1$, and how much more or less effective is $\theta_2$ relative to $\theta_1$. An illustration of the transformation approach is presented in \citet{ohagan:1998}.

Elicitation based on Gaussian copula allows the elicitation of a multivariate $p(\boldsymbol{\theta})$ to be decomposed into the elicitation of the marginals $p(\theta_i)$ and a so-called copula density. The method is based on the Sklar's theorem (1959) and the fact that the joint density $p(\boldsymbol{\theta})$ can be written as a product of the marginals $p(\theta_i)$ and the copula density, given that the marginal densities and the copula density are differentiable \citep{sklar:1959}. If the multivariate Gaussian copula is assumed with the density parameterized by a correlation matrix, then the elicitation reduces to elicitation of the marginals and the correlation matrix of the Gaussian copula. A detailed description of the method can be found in \citet{clemen:1999}; see also \citet{dallaglio:1991} and \citet{kurowicka:2006}. \citet{clemen:2000} compared six different dependence assessment methods (based on strength of relationship, correlation, conditional fractile, concordance probability, joint probability, and conditional probability) in two experimental studies. The main result was that directly asking the expert to report a correlation is a reasonable approach. The elicitation of correlations in the context of multivariate normal distribution has been studied by \citet{gokhale:1982} and \citet{dickey:1985}. 

One option is to directly elicit the joint or conditional probabilities of $p(\boldsymbol{\theta})$. \citet{fackler:1991} introduced a multivariate elicitation method based on quadrant probability or what he called the median deviation concordance (MDC) probability. MDC is defined as the probability that two variables will both fall either above or below their respective median. \citet{abbas:2010} proposed using isoprobability contours for multivariate elicitation. An isoprobability contour of a joint cumulative distribution function $F(\theta_1,...,\theta_d)$ is the collection of parameters $\boldsymbol{\theta}$ that have the same cumulative probability. They motivated the use of isoprobability contours by avoiding elicitation of the association or correlations between the parameters.

The elicitation of parameters of Bayesian neural networks (BNNs) serves as an extreme example of challenging multivariate elicitation. Models of this type can have millions of parameters without any interpretation attached to them. Prior elicitation for BNNs can be made possible by considering functional priors (e.g. Gaussian process priors, \citealp{sun:2019,tran2020all}) or informative weight space priors (e.g. sparsity-inducing Gaussian scale mixture priors, \citealp{louizos2017bayesian,fortuin2021bayesian}). An example of the latter is a prior elicitation method for BNNs introduced by \citet{cui:2021} that is based on eliciting summary statistics such as signal-to-noise ratio.

\subsection{Beta and Dirichlet prior}

The beta distribution typically describes the distribution of a proportion or a probability. It is the conjugate prior of the binomial sampling model in which case it describes the prior for the success probability. The elicitation of a beta prior has received considerable attention from the very beginning of the prior elicitation research. \citet{winkler:1967} essentially compared four elicitation techniques for a beta prior in a binomial sampling model. An overview of the elicitation methods for the beta distribution can be found in \citet{hughes:2002} with a summary in \citet[][Section 2.3.1]{jenkinson:2005}.

The methods differ with respect to which quantities are elicited and the procedure how the elicited values are converted into the parameters of a beta distribution. The methods of \citep{weiler:1965,fox:1966,gross:1971,holstein:1971,duran:1988,PHAMGIA:1992,enoe:2000} start by asking a location (mode, median or mean) of the beta distribution from the expert. A typical follow-up assessment task involves some dispersion measure or credible interval for the asked quantity. For instance, \citet{gross:1971} asked the mean $\mu$ and the expert's subjective probability that this mean lies in the interval $(0, K\mu)$, where $0<K<1$ is given by the analyst. In contrast, \citet{PHAMGIA:1992} elicited the median and the mean deviation about the median. It is also possible to first ask the credible interval, and then match the center of the interval with the mean of the beta distribution, as demonstrated by \citet{joseph:1995,cuevas:2012}. 

The methods of Equivalent Prior Samples (ESP) and Hypothetical Future Samples (HFS) \citep{good1965estimation,winkler:1967,bunn:1978} 
are model-specific (Hypercube D2) by assuming a binomial sampling model. Both methods first require the expert to assess the mean of the beta prior. Then, HFS requires the expert to update her prior mean given an imaginary sample, and ESP requires the expert to report the corresponding sample size. Both methods require expert feedback (mean of parameter) in the space of parameters (Hypercube D3), but EFS also asks a sample size, which is an observable quantity. So, EFS can be classified as a \textit{hybrid prior elicitation method}. Similarly, the beta prior elicitation methods introduced by \citet{chaloner:1983,gavasakar:1988} assume a binomial sampling model, being also model-specific. In contrast, the assessment tasks contain observables solely, such as asking a number of successes given a fixed number of trials. These types of methods are discussed in Section~\ref{observables_space_sec}.

\defcitealias{dorp:2004}{Van Dorp and Mazzuchi (2004)} 
Dirichlet distribution is a multivariate generalization of the beta distribution. Since the marginal distributions are betas, many elicitation methods for the Dirichlet distribution are based on univariate elicitations of the beta marginals. However, the problem is that the set of elicited beta marginal distributions would almost certainly not be consistent with any Dirichlet distribution. For instance, in order for the elicited beta marginals to be consistent with a Dirichlet distribution, the two parameters of a beta marginal have to sum to the same value for all beta marginals, and the expectations of the beta marginals have to sum to one. To overcome these issues, \citet{zapata:2014} and \citet{elfadaly:2017} suggest reconciling a suitable Dirichlet by specifying its parameters in terms of the parameters of ``corrected beta marginals". \citet{elfadaly:2013} proposed to use conditional corrected beta marginals. \citetalias{dorp:2004} studied how the parameters of beta marginals and Dirichlet extension can be recovered given a number of quantile constraints. Based on this work, \citet{srivastava:2019} considered elicitation of cumulative distribution function plots for the univariate marginal beta distributions for constructing a Dirichlet prior. \citet{evans:2017} let the expert state a lower or upper bound on each marginal probability that they are ``virtually certain" of.

The Dirichlet distribution is a conjugate prior of the categorical distribution and the multinomial distribution. Hence, the elicitation is often presented in the context of a multinomial model. For an overview, see \cite{wilson:2021} or \cite{Elfadaly:2012PhD}. Alternative approaches, which offer greater flexibility to the Dirichlet prior, are also introduced such as elicitation of generalized Dirichlet distribution \citep{elfadaly:2013}, mixture of Dirichlets \citep{regazzini:1999}, multivariate copulas \citep{elfadaly:2017} and vines \citep{wilson:2018}. The methods of \citet{dickey1983bayesian} and \citet{chaloner:1987} consider assessment tasks on the model observables (see Section~\ref{observables_space_sec}).

There are other grounds on which the parametric form of a prior can be based, other than conjugacy, which is often the case with the Beta and Dirichlet priors. The prior choice can be based on the maximum entropy principle \citep{jaynes1957information}, leading to \textit{maximal entropy priors}. Such priors are chosen to be the distribution with the largest entropy within the set of priors satisfying given constraints \citep{ZELLNER1977,Zellner1991,ZELLNER1996,bagchi:1986,Robert2007}. In our elicitation context, `the constraints' correspond to the elicited information from the expert. Abbas has studied both univariate \citep{abbas:2003} and multivariate \citep{abbas:2006} cases.

\subsection{Scoring rules}\label{scoring_rules}

Scoring rules are a class of devices for eliciting and evaluating probabilities. The concept appears in the classical work by \citet{savage:1971}, \citet{definetti:1974}, and \citet{murphy:1970}. The key idea behind scoring rule based elicitation is that the assessment tasks are formulated so that they encourage the expert to provide careful assessment from which subjective 'true' probabilistic judgment can be recovered. Indeed, in early work by \citet{brier:1950}, the idea of scoring was presented as a verification and 'bonus' system for forecasters to provide accurate predictions. \citet{matheson:1976} write: ``in terms of elicitation, the role of scoring rules is to encourage the assessor to make careful assessments and to be `honest', whereas in terms of evaluation, the role of scoring rules is to measure the `goodness' of the probabilities''. Hence, the scoring rule methods can be used either for elicitation or for evaluating the performance of an elicitation method. The focus in this subsection is on the former, whereas the latter is relevant for Main Section 3.4. See \citet{lindley:1987} for a philosophical justification and \citet{winkler1996scoring,gneiting:2007} for a review of scoring rules.

The scoring rules are derived from the expected utility hypothesis \citep{voneumann:1944}. The hypothesis states that an agent (expert) chooses between risky prospects or `lotteries' by selecting the one with the highest expected utility. For the sake of illustration, let the expert's subjective win probability be $p$, so her loss probability will be $1-p$. The analyst asks the expert to report her assessment for win probability, which is denoted by $q$. If the expert is risk-neural (explained later) and the analyst gives win reward $\log(q)$ and loss reward $\log(1-q)$ to the expert, then the expert reports her true subjective win probability, that is $q=p$. This result can be extended to continuous lotteries, where the expert reports a probability density $q(x)$ that reflects her subjective probability density $p(x)$ over possible outcomes $x$ \citep{matheson:1973,matheson:1976,holt:1979}. If we think of the parameter of interest $\theta$ as the outcome $x$, then the scoring rule methods can be applied to prior elicitation. The analyst formulates the elicitation of $p(\theta)$ as a gamble, where the expert bets on the plausible values of $\theta$ by choosing $q(\theta)$, and the rewards are designed so that the expert who maximizes the rewards chooses $q(\theta)$ close to $p(\theta)$.

The aforementioned term risk-neutral means that the underlying scoring rule implies that the expert is assumed to only care about the expected value of the payoff, not about how risky it is. In practice, it is difficult to choose the scoring rule that reliably reflects the expert's risk preference, and in principle that should be elicited also \citep{kadane:1988}. In our example, the analyst assumed a logarithmic scoring rule, but other scoring rules are also possible, such as quadratic and spherical \citep{holstein:1970,murphy:1970,matheson:1976}. \citet{karni:2009} and \citet{qu:2012} proposed a scoring based elicitation mechanism that allows incentives to be set at any desired level.

A drawback of scoring rule based elicitation systems is that they often require the expert to report a full probability distribution at once, which is a challenging task. This problem is mitigated in the ELI method \citep{VANLENTHE:1993,van:1993} that is a graphical computerized interactive elicitation technique, where scoring functions are displayed along with subjective probability density functions. The technique also graphically displays the consequences of a probability assessment, which resulted in a better calibration and a higher accuracy of the probability assessments in a study with 192 subjects \citep{VANLENTHE:1994}. The scores that allow elicitation by asking for quantiles instead of a full density function have been studied as well, see \citet{cervera:1996}, \citet{gneiting:2007}, \citet{schervish2012characterization}, and \citet{ehm:2016}. \citet{osband:1985}, \citet{lambert:2008}, \citet{gneiting:2011}, and \citet{steinwart2014elicitation} studied when a property of a probability distribution is \textit{elicitable} in scoring rule based elicitation. A property is elicitable if there exists a scoring function such that minimization of the associated risks recovers the property, or equivalently there exists a non-negative scoring function whose 'predictions that are further away from the property have a larger risk' \citep[Theorem 5,][]{steinwart2014elicitation}.

\section{Elicitation in observable space}\label{observables_space_sec}

Prior elicitation methods that let the expert provide her assessment in the space of model observables date back to the 1980s \citep{kadane:1980,kadane:1980b,winkler:1980}. The idea of deriving a prior for the parameters from a prior on the observables can be seen to be as old as Bayesian statistics itself \citep{stigler:1982}. In the prior elicitation literature, the idea of conducting the elicitation in the space of observables instead of parameters originates from the question about what is the essence of the model parameters, which is a controversial topic \citep{definetti:1937,geisser:1971,briggs:2016,billheimer:2019}. A natural requirement for a successful elicitation is that the expert is able to provide meaningful information on the asked quantities. When the asked quantity is a model parameter, the expert must know how to interpret it and have some idea of its natural scale and magnitude. In simple cases (e.g.\ the model is a Bernoulli trial) direct elicitation of the parameters can be justified on the grounds of the \textit{operational interpretation of the model parameters} in which the parameters are interpreted in terms of a (simple) limiting average of observables \citep{bernardo:1994}. Kadane et al.\ \citeyearpar{kadane:1980} took a position that:

\begin{quote} 
Without touching the philosophical issue of the sense in which parameters may be said to ``exist'', we acknowledge that even experienced observers have trouble answering questions on their beliefs about them.
\end{quote}
Later, \citet{kadane:1998} concluded that the choice of the elicitation space depends on the nature of the problem and whether the parameters have intrinsic meaning to the expert. For instance, applications in economics may benefit from elicitation methods that operate in the parameter space, since economist are ``quite used to thinking parametrically''. However, in general, \citet{kadane:1998} state that there has been some agreement in the statistical literature \citep[e.g.][]{winkler:1986,chaloner1996elicitation} on that the ``experts should be asked to assess only observable quantities''. The argument is that the expert does not need to understand what a parameter is. \citet{kadane:1998} also noted that the expert does not usually consider the correlation between regression coefficients, so elicitation in the observable space is particularly suited since it allows an indirect assessment of the correlations. \citet{wolfson:2001} argued that even though the methods in the observable space can be more difficult to implement, they can be more appropriate in terms of allowing the expert to communicate both her knowledge and uncertainty within the constraints of the model. \citet{akbarov:2009} argued that elicitation in the observable space can be made model-agnostic (Hypercube D2), since it need not rely on any assumptions about the family of distributions for the prior and sampling distributions, it can be designed to work with any prior predictive distribution. However, we found that there are relatively few model-agnostic prior elicitation methods that assume expert's feedback in the space of observables. For instance, we found one model-agnostic method \citep{hartmann:2020} compared to 27 model-specific methods in Table~\ref{table_model_specific}.

A challenge in doing elicitation in the observable space is the difficulty to separate two sources of randomness: due to $\theta$ and due to $y$ \citep{garthwaite:2005,perepolki:2021}. There are some solutions to that, for instance, \citet{kadane:1988} proposed to the expert to consider only the mean $\bar{y}$, and then conduct other assessment tasks to elicit the randomness in $y$. In linear regression, the latter amounts to eliciting the measurement error. \citet{stefan:2020} pointed out that obtaining unique prior distributions from the elicited data patterns on model observables becomes difficult for complex models, as for models with highly correlated parameters, the different prior parameter combinations can lead to similar model predictions on observables \citep{gutenkunst:2007}. For other arguments against and in favour of both elicitation schemes, see the discussed references and \citep{gaoini2009bayesian,choy:2009,james2010elicitator,al-labadi:2018}.

The main research line on elicitation in the observable space falls under the title of \textit{predictive elicitation} (see discussion in Hypercube D3). It assumes a setting where the expert is asked about the median, quantiles, mode, or expectation of the response variable (denoted by $y$) at various design points (denoted by $\mathbf{x}$, design matrix by $\mathbf{X}$), and the underlying model comes from the family of the generalized linear models (GLMs) \citep{kadane:1980,oman:1985,garthwaite:1988,ibrahim1994predictive,bedrick:1996,chen:2003,denham:2007,elfadaly:2011,garthwaite:2013,Elfadaly:2015}. The goal of the elicitation is to construct priors for the model hyperparameters. For a detailed review, see \citet{garthwaite:2005}; here we only discuss hyperparameters concerning regression coefficients. Subsequently, we discuss perhaps the most widely known method of this research line, the \textit{conditional means prior} (CMP) method \citep{bedrick:1996}. 

Given the model error variance $\sigma^2$, the regression coefficients are normally distributed with an unknown mean vector $\mathbf{b}$ and an unknown covariance matrix $\sigma^2 \mathbf{R}$. The elicitation of $\mathbf{b}$ can be as easy as setting $\mathbf{b} = (\mathbf{X}'\mathbf{X})^{-1}\mathbf{X}'\mathbf{y}_{.50}$ where $\mathbf{y}_{.50}$ is a vector of elicited medians \citep{kadane:1980}. However, the elicitation of $\mathbf{R}$ is difficult, since the matrix must be positive-definite and there can be many elements in the matrix. \citet{kadane:1980}, \citet{garthwaite:1988}, and \citep{garthwaite:2013} proposed an elicitation procedure where a structured set of sequential elicitation queries ensures that $\mathbf{R}$ is a positive-definite matrix. \citet{oman:1985} and \citet{ibrahim1994predictive} proposed setting $\mathbf{R} = c(\mathbf{X}'\mathbf{X})^{-1}$ where $c$ is a constant that is either estimated by using empirical Bayes or provided by the expert. \citet{oman:1985} argued that this approach is similar to a two-stage prior specification approach presented by \citet{schlaifer1961applied} and \citet{tiao1964bayes}, in that the first stage is a ``mental experiment'' where the design matrix $\mathbf{X}$ (which reflects the relations among explanatory variables) is specified by the expert. However, neither of the methods \citep{oman:1985,ibrahim1994predictive} uses the expert knowledge on the relationship between $y$ and $\mathbf{x}$ to infer $\mathbf{R}$, and thus the methods are not recommended for the elicitation of $\mathbf{R}$. In contrast, $\mathbf{R}$ is estimated from the elicited conditional percentiles of $y(\mathbf{x})$ in \citet{kadane:1980} and \citet{garthwaite:1988}

The CMP method  \citep{bedrick:1996} considers assessment tasks in which the expert specifies the mean responses corresponding to a collection of covariate combinations ($\mathbf{x}_1,...,\mathbf{x}_p$). By assuming the independence on the elicited mean responses, the induced prior distribution on the regression coefficients $\mathbf{b}$ can be derived. Specifically, given the GLM link function $g$, the collection of covariate combinations stacked in the matrix $\Tilde{\mathbf{X}} \in \mathbb{R}^{p \times p}$, and the elicited conditional means $m_i = \mathrm{E}(y_i|\mathbf{x}_i)$ with $\mathbf{m}=(m_1,...,m_p)$, the prior can be obtained from $\mathbf{b} = \Tilde{\mathbf{X}}^{-1}g(\mathbf{m})$ by using the transformation formula. The method is widely adopted with examples found in clinical trial design \citep{yuan2016bayesian}, ecology \citep{kynn2006designing}, medicine \citep{zeng2011bayesian}, and statistics \citep{gelman:2008,pires:2019}. There is an interesting connection to the Zellener's \citeyearpar{zellner1983applications} g-prior, namely \citet{hanson:2014} showed that the g-prior is a particular conditional means prior with $\Tilde{\mathbf{X}}$ chosen to be “canonical covariates".

Complementing the main research line on elicitation in the observable space, \citet{garthwaite:1992}, \citet{laud:1995}, and \citet{chen:1999} studied feature selection in regression models \citep[see also][in a context of power priors]{ibrahim:2000,ibrahim:2015}. The elicitation method of \citet{garthwaite:1992} is hybrid in that it combines elicitation data from both the parameter and observable spaces \citep[see also][]{good1965estimation,denham:2007,casement:2018}. Another elicitation problem that has got some attention is the prior elicitation of a multivariate normal sampling model. Al-Awadhi et al.\ \citeyearpar{al-awadhi:1997,al-awadhi:1998} assumed the natural conjugate prior (normal inverse-Wishart) and \citet{garthwaite:2001} assumed a non-conjugate prior (normal generalized inverse-Wishart). Natural conjugate prior forces a dependence between the mean and the covariance, so \citet{garthwaite:2001} proposed assessment tasks that allow the expert to quantify separately assessments about each of these parameters. Assessment tasks include conditional and unconditional quantiles where the conditions were specified by hypothetical data. \citet{al-awadhi:2001} compared the approaches of \citet{al-awadhi:1998} to \citet{garthwaite:2001}, and concluded that the independence between mean vector and covariance matrix \citep[which holds in][]{garthwaite:2001} leads to better performance in terms of obtained scoring rule values. Furthermore, there are prior elicitation methods for special applications, where the expert's assessment task contains assessing observables, such as in survival analysis \citep{coolen:1992,wang:2009}, risk control \citep{hosack:2017}, item response theory \citep{tsutakawa1984estimation}, and flash flood prediction \citep{gaoini2009bayesian}. 

Many authors have pointed out that the \textit{prior predictive distribution},
\begin{equation}\label{priorpredictive}
    p(y) = \int p(y|\theta)p(\theta)d\theta,
\end{equation}
gives rise to an integral equation when the distributions concerning the observables $y$ are known \citep{aitchison:1975,winkler:1980,wolpert:2003,gribok:2004,akbarov:2009,jarocinski:2019}. Suppose that $p(y)$ is elicited from the expert, the likelihood $p(y|\theta)$ is specified by the analyst, and the analyst is looking for the expert's prior $p(\theta)$. Then, the equation \eqref{priorpredictive} is known as a Fredholm integral equation of the first kind, which is a well-known example of linear ill-posed problems. Additional regularity assumptions are needed in order to solve the prior from this equation.  For instance, \citet{gribok:2004} proposed Tikhonov regularization that imposes smoothness constrains on the prior density, which is a natural restriction because probability density functions are typically smooth and differentiable. Alternatively, the analyst may assume a parametric prior $p(\theta|\lambda)$ where $\lambda$ are its hyperparameters, in which case the prior predictive distribution reads as $p(y|\lambda) = \int p(y|\theta)p(\theta|\lambda)d\theta$, and the problem reduces to finding the optimal hyperparameters. For instance, Percy \citeyearpar{percy2002bayesian,percy2003subjective,percy2004subjective} considered beta, gamma and normal priors, and demonstrated how the problem reduces to solving a system of nonlinear equations. \citet{yuan2016bayesian} considered a univariate logistic regression and approximated the integral equation so that $\lambda$ can be estimated from the elicited prior mean probabilities for $y=1$ at two covariate locations.

The prior predictive distribution provides a link between the model observables and the parameters, and the aforementioned integral equation is only one way to use that link in prior elicitation. In a recent work, \citet{hartmann:2020} assumed that the expert's assessment task consists of assessing (prior predictive) probabilities of observables falling in certain regions of the observables space. The elicited probabilities were treated as noisy data with a suitable likelihood (see supra-Bayesian approach in Hypercube D4). The prior predictive distribution can also be used in prior elicitation via computer simulations. Monte Carlo integration of the integral \eqref{priorpredictive} involves sampling first $\theta' \sim p(\theta)$ and then $y' \sim p(y|\theta')$.  
\citet{wesner:2021} considered comparing $y'$ to some reference value $y^*$, and then back-tracking $\theta'$ that produced $y'$ close to $y^*$. \citet{thomas2020probabilistic} considered two assessment tasks: verisimilitude judgements (is $y'$ a credible draw from reality?) and pairwise judgements (for given $y'$ and $y''$, which one is more realistic?). The authors trained a Gaussian process classifier on this binary label data to recover the expert's prior distribution on realistic parameters. \citet{chandramouli2016extending} and \citet{chandramouli2020accommodation} proposed a framework where ``data priors'' are specified over the space of predictive distributions. Priors on model parametrizations are informed by how well their predictions approximate the different predictives.

\section{Nonparametric priors}\label{nonparametric_prior}

How should the functional form of prior distribution be chosen? Given that the goal of prior elicitation is to faithfully represent the belief of the person being elicited \citep{garthwaite:2005,ohagan:2006}, the family of parametric prior distribution or the model of nonparametric prior should be chosen to reflect this belief. However, also other objectives and practical aspects influence the choice of prior family, for example a specific parametric prior may be easier to use with probabilistic inference engines. For instance, the natural conjugate family is often chosen for computational ease \citep{garthwaite:2005,BDA:2013}, although also `piecewise conjugate priors' \citep{meeden1992elicitation} and `mixture of natural conjugate priors' \citep{dalal1983approximating} have been proposed. Even if we agree that the choice of prior should be based on available information, and not for instance computational convenience, there is still some ambiguity in specifying the prior family. \textit{Conservative methods} have been designed to avoid arbitrary choices among all possible prior distributions whose features can fit with the elicited information from the expert \citep{Robert2007,press:2002,bousquet:2008}. On the other hand, \textit{non-conservative methods} incorporate additional information into the prior by selecting a functional form for the prior that reflects assumptions about the expert's knowledge distribution. For instance, \citet{ohagan:2007} introduced a nonparametric supra-Bayesian elicitation framework where the expert's knowledge (a Gaussian process model) is assumed to have a specific prior mean function that differs from the zero-mean function.

Elicitation of a parametric prior has a limitation: it forces the expert's knowledge distribution to fit the parametric family \citep{ohagan:2007}. Even though it is not reasonable to assume that there exists a `true' prior distribution waiting to be elicited but rather a ‘satisfying’ prior distribution \citep{winkler:1967}, the fitted parameteric prior may fall too far away from a distribution that reflects the expert's knowledge distribution `satisfactorily' well. A nonparametric prior, such as a histogram or a Gaussian process, can capture a wider range of elicitation data structures than parametric priors, and thus provides more flexibility for finding a `satisfying’ prior.

Some early prior elicitation methods can be regarded as nonparametric, such as \citep{schlaifer:1969}, where the expert is asked to directly draw the density function. Similarly, the histogram method can be used for eliciting a nonparametric prior distribution \citep{berger1985}. However, more complex nonparametric prior models can be found in the literature. \citet{ohagan:2007} proposed modelling the expert's density, $p(\theta)$, as a Gaussian process model (GP) that is completely characterized by its mean and covariance function \citep{williams2006gaussian}. Since this approach treats the expert's prior density as a random function, the analyst can, for example, look at the expectation of the GP model to obtain the resulting (deterministic) prior. The choice of the mean and covariance function provides a way to incorporate assumptions about the expert's knowledge distribution. For instance, the authors proposed a covariance function that reflects the assumption that for those parameter values where the expert's density is small, the prior GP variance should be small there too. The assessment tasks involve the expectation and percentiles of the distribution of $\theta$. These elicited summaries are then used for inferring the posterior distribution for the GP model.

\citet{gosling:2007} developed the method of \citet{ohagan:2007} further. The authors raised two deficiencies in the method. First, they pointed out that the analyst’s uncertainty about the tails of the expert’s knowledge density can be underestimated (recall that the analyst models the expert's density as a random function). Second, there are situations where the method assumes that the analyst knows the expert's true density with certainty. The authors tackled these challenges by assuming the prior mean of the expert's knowledge distribution to follow the Student's $t$-distribution. \citet{oakley:2010b} and \citet{daneshkhah:2006} coupled the method of \citet{ohagan:2007} with the roulette method (see Section~\ref{Quantiles,moments...}). This reduces the cognitive burden of the expert, however, at the expense of the method becoming more computationally intensive. \citet{daneshkhah:2006} also investigated the issue of imprecision in the expert probability assessments in \citet{ohagan:2007}. The authors modelled the imprecision in the expert’s probability assessments by assuming that the elicited probabilities are contaminated by additive noise (Gaussian or uniform). They concluded that this ``has sensible influence on the posterior uncertainty and so the expert density''. An example of a nonparametric elicitation method that does not use GPs is provided by \citet{gertrudes:2001}. The authors represented the prior as a convex combination of distributions fitting the expert's data.

\citet{moala:2010} proposed a multivariate extension of \citet{ohagan:2007}. In multivariate setting, the probabilities of the form $P(\boldsymbol{\theta} \in A)$, where $A$ is a region of the parameter space, should be considered instead of percentiles. The authors illustrated that good results can be obtained by eliciting marginal probabilities supplemented by a small number of joint probabilities. However, the authors considered only bivariate elicitation, and they did not demonstrate that the method can deal with a higher number of parameters than two. The expert's nonparametric (GP) model was capable of adapting to a bimodal ground-truth distribution (in a numerical experiment), showing that this nonparametric method is flexible enough to capture a target that is not a simple unimodal distribution. \citet{hahn:2006} introduced a nonparametric elicitation method that can be applied to multivariate elicitation. The method essentially constructs a histogram from the expert's data that consists of the elicited relative odds of parameters. \citet{langlois:2021} elicited prior distributions on locations in (2D) visual scenes by modelling the priors as Gaussian kernel density estimators.

\section{Model-specific methods}\label{model_specific_methods}

We will not address model-specific elicitation methods in detail, since we consider it more relevant to focus on model-agnostic methods such as \citep{gelfandetal:1995,gosling:2005,ohagan:2007,oakleyshelf:2019,hartmann:2020} discussed in the other parts of the article. However, some model-specific methods have been designed for model families that still offer broad flexibility in terms of practical applications, and hence can be valuable in practical modelling tasks. An excellent example is the research line on generalized linear models (GLMs) started by \citet{kadane:1980} and \citet{bedrick:1996} (more about that in Section~\ref{observables_space_sec}). While GLMs are clearly a limited model family, they are widely used in several fields and good elicitation tools for them can be useful. 
We refer the reader to Table~\ref{table_model_specific} that categorizes the model-specific methods according to the model, the prior and the elicitation space, as a starting point for investigating the literature on model-specific elicitation methods for different cases.

\noindent\fbox{%
    \parbox{\textwidth}{
\underline{The abbreviations in Table~\ref{table_model_specific}.} NLR = normal linear regression, GLM = generalized linear model, LR = logistic regression, MVN = multivariate normal, Dir = Dirichlet, Bin = Binomial, Mult = Multinomial, Gam = Gamma, Poiss = Poisson, BQR = Bayesian quantile regression, AR = autoregressive model, PHR = proportional hazards regression, BN = Bayesian network, CIHM = conditionally independent hierarchical model, H = hierarchical, RE = random effect, PW = piecewise, (N)CP = (natural) conjugate prior, GIW = generalized inverse Wishart, PC = penalized complexity prior \citep{simpson:2017}, CMP = conditional mean prior \citep{bedrick:1996}, O = observable space, P = parameter space, H = hybrid space
    }
}

\defcitealias{oakleyshelf:2019}{SHELF}
\begin{table*}
\begin{tabular}{llllH}
\hline
\textbf{Author} & \multicolumn{1}{l}{\textbf{Model}} & \multicolumn{1}{l}{\textbf{Prior}} & \multicolumn{1}{l}{\textbf{Hypercube D3}} &  \textbf{Notes} \\
\hline
\citet{kadane:1980} & NLR & NCP & O &  \\ \hline 
\citet{oman:1985} & NLR & NCP & O &  \\ \hline
\citet{garthwaite:1988} & NLR & NCP & O  &  \\ \hline
\citet{ibrahim1994predictive} & NLR & NCP & O &  \\ \hline
\citet{bedrick:1996} & GLM & CMP & O &  \\ \hline
\citet{chen:2003} & GLM & CP & O &  \\ \hline
\citet{denham:2007} & GLM & NCP & H & geographic applications \\ \hline
\citet{elfadaly:2011} & NLR & NCP & O &  variance elicitation \\ \hline
\citet{garthwaite:2013} & PW-GLM & NCP & O &  \\ \hline
\citet{Elfadaly:2015} & Gam-GLM & NCP$/$log-normal & O &  extra parameters elicitation\\ \hline
\citet{garthwaite:1992} & NLR & mixture-NCP & H & feature selection \\ \hline
\citet{laud:1995} & NLR & NCP & O & feature selection \\ \hline
\citet{chen:1999} & LR  & custom & O & historical, feature selection \\ \hline
\citet{leamer:1992} & NLR & NCP & P &  variance elicitation\\ \hline
\citet{hosack:2017} & GLM & NCP & O &  \\ \hline
\citet{carlin:1992} & RE-LR & custom & P & application+historical\\ \hline
\citet{alhamzawi:2011} & RE-BQR  & power prior & P &  application \\ \hline
\citet{garthwaite:2006} & PW-LR  & NCP & O &  \\ \hline
\citet{kadane:1996} & AR  & PW-CP & O &  \\ \hline
\citet{garthwaite:1991} & NLR & NCP & O &  \\ \hline
\citet{chaloner:1993} & PHR & adjusted-NCP & O & clinical application \\ \hline
\citet{ibrahim:1999} & PHR & semi-parametric & O &  \\ \hline
\citet{soare:2016} & NLR & delta & P & feature selection \\ \hline
\citet{micallef:2017} & NLR & half-normal & P & feature selection \\ \hline
\citet{afrabandpey:2017} & NLR & NCP & O & feature selection \\ \hline
\citet{hem2021} & RE-H  & PC/Dir & P & variance elicitation \\ \hline
\citet{rizzo:2019} & BN  & nodepriors & P & \citetalias{oakleyshelf:2019} based\\ \hline
\citet{alawadhi:2006} & PW-LR  & NCP+other & O & ecology application \\ \hline
\citet{kynn:2005} & PW-LR  & CMP & O & ecology application \\ \hline
\citeauthor{oleary:2008} \citeyearpar{oleary:2008,oleary:2009} & LR  & mixture-normal & P &  \\ \hline
\citet{al-awadhi:1998} & MVN  & NCP & O &  \\ \hline
\citet{garthwaite:2001} & MVN  & GIW & O &  \\ \hline
\citet{steffey:1992} & CIHM  & custom & P &  \\ \hline
\citet{elfadaly:2020} & Mult-LR  & logit-normal & O &  \\ \hline
\citet{bunn:1978} & Mult  & Dir & P & \\ \hline 
\citet{chaloner:1987} & Mult  & Dir & O &  \\ \hline
\citet{dickey1983bayesian} & Mult  & Dir & H & obs in re-assessment stage  \\ \hline
\citet{regazzini:1999} & Mult  & mixture-Dir & O &  \\ \hline
\citet{wilson:2018} & Mult  & D-vine & P &  \\ \hline
\citet{elfadaly:2013} & Mult  & Dir/Connor-Mosimann & P &  \\ \hline
\citet{elfadaly:2017} & Mult  & Dir/Gaussian copula & P &  \\ \hline
\citet{srivastava:2019} & Mult  & Dir & P &  \\ \hline
\citet{moreno:1998} & Poiss-Bin  & NCP & P &  \\ \hline
\citet{good1965estimation} & Bin  & Beta & P/H &  ESP and HFS methods \\ \hline
\citet{chaloner:1983} & Bin  & Beta & O &  \\ \hline
\citet{gavasakar:1988} & Bin  & Beta & O &  \\ \hline
\end{tabular}
\caption{Summary of model-specific elicitation methods.}
\label{table_model_specific}
\end{table*}

\section{Active elicitation}\label{active_elicitation}

Human effort is costly, so an elicitation algorithm should assume asking the expert on a budget. Active elicitation helps to make most out of this limited budget, and it is relevant when overfitting (see Hypercube D4) is considered. Active learning \citep{cohn:1994,settles:2012,gal:2017} and Bayesian experimental design \citep{chaloner:1995,ryan:2016} study the selection of informative data-points (or `experiments') to be used in regression or classification tasks. When active learning is applied to prior elicitation, we call it active prior elicitation or just \textit{active elicitation}. While active learning typically chooses new data to maximize information gain about the parameters, active elicitation can use the same principles, but instead querying new data from the expert.

Various strategies have been proposed for active prior knowledge elicitation of the weights of a sparse linear regression model. For instance, \citet{daee:2017} used a strategy that maximizes the expected information gain in the prediction of an outcome (see \citealp{seeger:2008} for a non-elicitation context). The higher the gain, the greater impact the new expert's input has on the predictive distribution. \citet{sundin:2018} studied this gain in a personalized medicine scenario where a single prediction was considered instead of predicting over a whole training dataset. \citet{micallef:2017} modelled the expert's binary relevance input on the regression coefficients as a linear bandit algorithm. The authors proposed to select the next query by maximizing a high probability bound for the uncertainty of feature relevance.

Active elicitation assumes iterative elicitation (see Hypercube D6). This enables the theoretical study of the algorithmic convergence in elicitation tasks. However, surprisingly, we are not aware of any results of convergence guarantees in prior elicitation tasks. For instance, in clustering tasks, \citet{balcan2008clustering} and \citet{awasthi:2014} provided theoretical guarantees for the convergence of a clustering algorithm where the expert can be queried about split and merge requests. Similarly, \citet{asthiani:2016} provided convergence guarantees when the expert can be queried about instances belonging to same clusters. \citet{kane:2017} extended the theoretical study to comparison queries.

In active elicitation, modelling of the expert's behaviour becomes important. \citet{daee:2018} showed a significant performance improvement of using a user model, to model the expert's behaviour, in a feature relevance elicitation task for sparse linear regression. The expert is not treated as a passive data source but instead her behaviour is modelled to anticipate her input and take into account her biases, see also \citet{micallef:2017,afrabandpey:2019}. Active elicitation could be used to minimize experts’ effort in any type of prior elicitation, in principle. In many works cited here, the expert was shown data during the elicitation, which is an unconventional choice in prior elicitation. The motivation was to further reduce the expert’s effort. Furthermore, in some works, the elicitation data was used as separate auxiliary data.

\section{Multiple experts}\label{multiple_experts}

Eliciting the knowledge of a group of experts is an important research line. However, in this article, we do not go into this topic in any great depth, but instead we will confine ourselves to providing the key references on the subject. A brief summary of the topic can be found in \citeauthor{ohagan:2019} \citeyearpar[][Sec. 3]{ohagan:2019}, \citeauthor{garthwaite:2005} \citeyearpar[][Sec. 5]{garthwaite:2005}, and \citeauthor{jenkinson:2005} \citeyearpar[][Sec. 7]{jenkinson:2005}. We refer the reader to the following papers \citep{winkler:1968,albert:2012,hanea:2018,phillips:1999,phillips:1993,french:2011,hartley2018elicitation,ouchi:2004,daneshkhah:2011,genest:1985,morris:1977}, and to a recent comparison study by \citet{williams:2021}.

Around the elicitation of a group of experts, there have been research on protocols how to coordinate the elicitation process. Four prominent elicitation protocols have been designed for elicitation with multiple experts: Sheffield protocol \citep{gosling:2018,oakleyshelf:2019}, Cooke protocol \citep{cooke:91}, Delphi protocol  \citep{brown:1968}, and IDEA protocol  \citep{hanea:2018}. An overview of the protocols is provided by \citet{european:2014}, \citet{dias:2018}, and \citet{ohagan:2019}. 

Finding a single distribution that combines elicited knowledge from all the experts is known as the problem of aggregation. There are two main approaches. In the \textit{mathematical aggregation}, the experts are elicited separately, and
a probability distribution is fitted to each expert’s knowledge. These are then combined into an aggregate distribution using a mathematical formula, a so-called \textit{pooling rule}. A popular pooling rule is \textit{supra-Bayesian pooling} \citep{morris:1974,morris:1977,keeney:1976,lindley1985reconciliation,west:1988,genest1986combining,roback:2001,albert:2012}. The \textit{behavioural aggregation} encourages the group of experts to discuss their knowledge, and to settle upon group consensus judgments, to which an aggregate `consensus' distribution is fitted.

\section{Heuristics and biases}\label{heuristics_biases}

Every expert is inclined to commit to cognitive biases and to use cognitive shortcuts (heuristics) when making probabilistic judgments. Tversky's and Kahneman's heuristics and biases research program in the 1970s led to a series of published papers on describing difficulties in assessing probabilities. The work was summarized in the seminal paper by \citet{tversky:1974}. \citet{ohagan:2019} highlighted four relevant heuristics and biases for expert knowledge elicitation: anchoring, availability, range-frequency, and overconfidence, of which the first two already appear in the Tversky's and Kahneman's original article \citeyearpar{tversky:1974}. We encourage the reader to take a look at an interesting summary of these provide by O'Hagan \citeyearpar{ohagan:2019}, although there has also been some controversy on the topic \citep{gigerenzer1996narrow,kynn:2008}, such as an allegation that the statistical literature ignores the effect of framing in elicitation.

Every prior elicitation method should address systematic cognitive biases identified in the literature. Indeed, the leading principle in designing the elicitation protocols mentioned in Section~\ref{multiple_experts} has been to minimize cognitive biases to which expert probabilistic judgments may be subject \citep{ohagan:2019}.

We provide some entry-points to this topic. We refer the reader to \citeauthor{ohagan:2019} \citeyearpar[][Sec. 2]{ohagan:2019}, \citeauthor{garthwaite:2005} \citeyearpar[][Sec. 2]{garthwaite:2005}, \citet{kynn:2008}, \citet{gigerenzer:1999}, and \citet{burgman:2016}. See also the discussion on how to deal with imprecision in probability judgements \citep{ohagan:2004}.

\bibliographystyle{ba}
\bibliography{refs}